\documentclass[prb,twocolumn,amsmath,amssymb,showpacs]{revtex4}

\usepackage{graphicx}
\usepackage[latin1]{inputenc}

\begin{document}

  \title{
    Electronic transport calculations for self-assembled mono-layers of\\
    1,4-phenylene diisocyanide on Au(111) contacts
  }
  \author{
    Robert Dahlke
    and Ulrich Schollw\"{o}ck 
  } 
  \affiliation{
    Sektion Physik and Center for Nanoscience, LMU M\"{u}nchen, 
    Theresienstrasse 37, 80333 M\"{u}nchen, Germany
  } 
  \date{October 8, 2003}

  \begin{abstract}
    We report on electronic transport calculations for self-assembled
    mono-layers (SAM) of 1,4-phenylene diisocyanide on Au(111)
    contacts.  Experimentally one observes more structure (i.e\ peaks)
    within the measured conductance curve for this molecule with two
    cyanide end-groups, compared to measurements with molecules having
    thiol end-groups.  The calculations are performed on the
    semi-empiric extended H\"{u}ckel level using elastic scattering
    quantum chemistry (ESQC) and we investigate three possible
    explanations for the experimental findings.  Comparing the
    experimental and theoretical data, we are able to rule out all but
    one of the scenarios. The observed additional peaks are found to
    be only reproduced by a mono-layer with additional molecules
    perturbing the periodicity.  It is conjectured that the weaker
    coupling to Au of cyanide end-groups compared to thiol end-groups
    might be responsible for such perturbations.
    
  \end{abstract}
  \maketitle

  \section{Introduction}\label{sec:introduction}
  Within the last decade an increasing interest in molecular
  electronics has developed, with the expectation of realising
  molecular diodes and transistors.  This is based on the progress in
  manipulation techniques, which now allow the controlled attachment
  of atomic scale structures like molecules to mesoscopic leads.  With
  these new devices one is able to determine the conductance
  properties of molecular structures.  Explaining and predicting the
  electronic behaviour of such devices is an essential step towards
  their design and use as nano-scale electronic circuits.
  
  To this end a number of theoretical studies have been performed with
  the aim of reproducing measured $IV$ characteristics.  These studies
  differ in the way they take the electronic levels of the molecules,
  their modification by the coupling to the leads and the change of
  electrostatic potential due to bias into account.  Semi-empiric
  methods\cite{Datta1997:PRL,Magoga1997:PRB,Emberly2001:PRB} have been
  used, as well as first principles
  techniques,\cite{Derosa2001:JPhysChemB,Damle2002:ChemPhys,Brandbyge2002:PRB,Krstic2002:PRB}
  the latter being restricted to systems of moderate size.
  
  The wide range of experimentally observed behaviour (see section
  \ref{sec:experiment}) suggests, that not only the structure of the
  molecule, but also the details of the device fabrication affect the
  conduction properties of molecular devices.  The crucial step is the
  deposition of molecules onto the surface of the lead. As this is
  achieved by self-assembly the amount of adsorbed molecules and their
  individual positions can not be exactly controlled and therefore
  remains unknown.  A satisfactory understanding of the interplay
  between geometrical alignment of the molecules and measured
  conductance properties has thus not yet been achieved (for a recent
  review see e.g.\ Nitzan et al.\cite{Nitzan2003:Science}).
  
  In this paper we study the way in which changing the geometrical
  alignment of the mono-layer has an influence on the conduction
  properties of a molecular device.  In so doing we can rule out a
  number of explanations which have previously been
  considered\cite{Dupraz2003:unpublished} to explain the occurrence of
  additional structure in the conductance-voltage ($CV$)
  characteristics.
  
  The outline is the following: first we summarise some of the recent
  experimental findings. Then the method we use (based on
  ESQC\cite{Sautet1989:CPL}) for calculating the conductance
  properties of molecular devices, is discussed.  Calculations for the
  conductance of a SAM, being build of 1,4-phenylene diisocyanide
  (PDI) and sandwiched between gold leads are then presented.
  The results for three
  qualitatively different geometrical constitutions of the mono-layer
  are compared with experimental data.
  By this we can rule out all but one and conclude that
  the only geometrical alignment, which gives rise to several peaks in
  the conductance curve, is a mono-layer with additional molecules
  perturbing the periodicity.
  
  \section{Experimental overview\label{sec:experiment}}
  The devices built to study conductance properties of molecular
  structures differ not only in amount and chemical structure of the
  molecules in use but also in the way these are attached to metallic
  or semi-conducting leads.  Single or few molecules are accessible in
  mechanically controllable break junctions (MCB) and with the
  scanning tunneling microscope (STM).  Many molecules are involved
  in sandwiched self-assembled mono-layer (SAM) experiments.  The
  observed properties depend on the exact geometry of the device.  The
  conductance differs in orders of magnitude and the qualitative
  voltage dependency of the current ranges from simple ohmic behaviour
  to negative differential resistance (NDR).\cite{Chen1999:Science}
  
  In the past Reed et al.\cite{Reed1997:Science} have measured the
  electrical conductance of a self-assembled molecular mono-layer
  bridging an MCB at room temperature.  Molecules of
  1,4-benzene dithiol (i.e.\ having two thiol end-groups, which are
  known to couple strongly to Au-atoms) were used and the $CV$
  characteristic was found to be symmetric with one peak in the
  voltage range of $0-2$V.  They measured a current of the order of
  $50$nA at a bias voltage of $2$V, which they claim is produced by
  transport through one single active molecule.  Reichert et
  al.\cite{Reichert2002:PRL} also used an MCB with molecules having
  two thiol end groups, but being considerably longer.  The measured
  current amplitude was about $500$nA at $1$V, i.e.\ although the
  molecule was more than twice as long, the current was ten times
  larger.
  
  With a different setup, where a SAM is sandwiched between two
  metallic leads, Chen et al.\cite{Chen1999:Science} have found
  negative differential resistance (NDR), namely one peak at $2$V in
  the $IV$ curve.  The molecule under investigation had one thiol
  end-group only and was attached to Au-leads at both ends.  The
  measurements were taken at room temperature and the measured current
  maximum was of the order of $1$nA.
  
  Only recently, sandwiched SAM devices at $4.2$K were
  studied,\cite{Lee2003:NanoLet,Dupraz2003:unpublished} where a
  benzene ring with two cyanide instead of thiol end-groups was used.
  The measurements exhibited currents of the order of $50-400$nA. The
  $CV$ characteristic for this molecule revealed more structure, in
  form of three to five peaks within a voltage range of $1$V.  Such a
  behaviour was not observed with previous devices containing other
  molecules.
  
\section{Theoretical Formalism}
  In the literature there has been presented
  quite a number of techniques
  to calculate non-equilibrium
  electronic transport through molecular systems attached to
  mesoscopic leads.
  Usually the Landauer formalism is applied, which
  describes current as elastic electron transmission and
  therefore requires the transmission function $T(E)$.
  To this end one needs a framework that allows for a
  description of the molecular device
  on the atomic level. This involves not only
  the molecules themselves, but also the surface and bulk region of
  the leads. Quantum chemistry provides such a framework, and
  one can choose the level of theory according to the size of the
  system under consideration and the computational effort one is willing
  to spend.
  
  Using a quantum chemistry method, 
  the transmission function can be obtained from an 
  effective one-particle Hamiltonian, which is
  an appropriate description for strong coupling of the molecules
  to the leads 
  (as in the case of covalent binding).
  The methods differ in the generation of the one-particle Hamiltonian,
  which might be based on semi-empirical grounds
  \cite{Datta1997:PRL,Magoga1997:PRB,Emberly2001:PRB}
  or on first-principles and self-consistent techniques.
  \cite{Derosa2001:JPhysChemB,Damle2002:ChemPhys,Brandbyge2002:PRB,
  Krstic2002:PRB}
  
  A different approach,\cite{Hettler2003:PRL} 
  taking many-particle effects explicitly into
  account, uses a master equation with transition rates calculated
  perturbatively using the golden rule.  
  This approach is
  appropriate for weak coupling.
  
  We use the Landauer formalism, as the molecules are assumed to be
  chemically bonded to the gold contacts (i.e.\ strong coupling),
  together with the semi-empiric extended H\"{u}ckel quantum-chemistry
  method ESQC.\cite{Sautet1989:CPL} The molecular structure is
  optimised\cite{Gaussian98} beforehand.  This approach, though
  limited as compared to more sophisticated quantum chemistry methods,
  is yet justified, because we want to gain a qualitative
  understanding of a many molecule experiment, which can not be
  described by first-principle techniques, as the number of atoms
  involved is beyond the practical limitations of to-date computer
  resources.
  
  \subsection{Landauer formalism}
  According to the Landauer formula, current along a defect region is
  the result of electron transmission from the source to the drain
  lead.  This is described by the transmission function $T(E)$ and net
  electron transfer happens at all those energies, where there are
  more states occupied in the source lead than in the drain.  The
  difference in occupation is a result of the applied bias $V$, which
  raises the chemical potential $\mu_1$ of the source lead above the
  one of the drain lead $\mu_2 = \mu_1 - eV$ and thus changes the
  distribution function $f_i(E)=f(E-\mu_i)$.
  \begin{equation}
    I = \frac{-2e}{h}\int_{-\infty}^\infty T(E)\left(f(E-\mu_1)-f(E-\mu_2)\right) dE.
  \end{equation}
  The Landauer formula is valid under the condition that transport is
  coherent across the molecule, which is plausible as the typical
  mean-free path of electrons within metals is of the order of
  $500$nm, while the molecular gap between source and drain lead is
  only about $1-5$nm in length.  Furthermore it is assumed that the
  coupling of the leads to macroscopic reservoirs is reflection-less,
  and that the tunneling electrons equilibrate only deep inside the
  leads.  This ensures that the distribution function for incoming
  electrons in a lead can be taken to be spatially constant, even
  close to the molecular region.
  
  \begin{figure}
    \includegraphics[height=0.5\textwidth,angle=270]
    {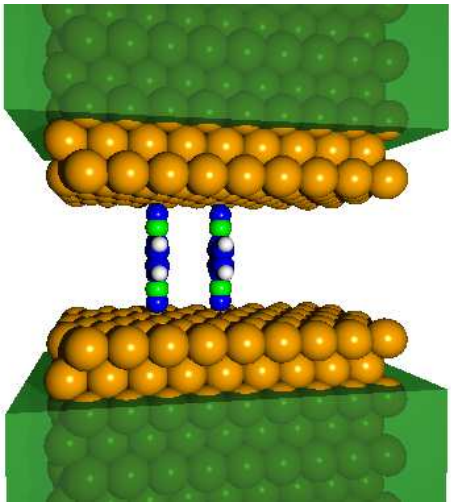}
    \caption{\label{fig:system}(Colour online) Partitioning of the system into three parts: 
      the two semi-infinite leads $\Sigma_{1,2}$ (surrounded by boxes) 
      and the molecular region $\Sigma_0$ between them.}
    \label{fig:square_well}
  \end{figure}
  
  The system is formally partitioned into three regions $\Sigma_i,
  i\in\{0,1,2\}$, two of them ($\Sigma_{1,2}$) containing the
  semi-infinite leads, the third one ($\Sigma_{0}$) being the finite
  region containing all molecules as well as a few surface layers of
  each lead (see Fig.\ \ref{fig:system}). We use periodic boundary
  conditions in the directions perpendicular to the surface normal.
  
  By a tight binding approximation, the infinite-dimensional
  Hamiltonian of the entire system can be 
  composed of quantum chemical 
  one-particle block Hamiltonians of finite dimension:
  \begin{align}\label{eq:SystemHamiltonian}
    \nonumber
    H 
    &=
       \sum_{i\in\text{mol}}(\varepsilon^{\phantom x}_{i} 
         c^\dagger_{mi}c^{\phantom x}_{mi}
         +\sum_{j\neq i}
         H^m_{ij}c^\dagger_{mi}c^{\phantom x}_{mj})
\\\nonumber
    &+ 
       \sum_{d\in\text{leads}}\sum_{l=l_0}^\infty \sum_{i\in l} 
         (\varepsilon^{\phantom x}_{dli}c^\dagger_{dli}c^{\phantom x}_{dli}+
         \sum_{j\neq i}H^d_{llij}c^\dagger_{dli}c^{\phantom x}_{dlj})
\\\nonumber
    &+ 
       \sum_{d\in\text{leads}}\sum_{l=l_0}^\infty
         \sum_{\substack{i\in l\\j\in l+1}}(
         H^d_{l,l+1,ij}c^\dagger_{dli}c^{\phantom x}_{d,l+1,j}
         +\text{h.c.})
\\
    &+
       \sum_{d\in\text{leads}}\sum_{i\in l_0}\sum_{j\in\text{mol}}(H^{dm}_{l_0ij}
         c^\dagger_{dl_0i}c^{\phantom x}_{mj} +\text{h.c.})
  \end{align}
  The first summation describes the isolated molecular region,
  by an on-site energy $\varepsilon$ and a hopping term. 
  The indices $i$ and $j$ run
  over the orbital basis set within that region.
  The next two summations describe the isolated leads,
  labelled by $d$. 
  Layer by layer, starting with the surface layer $l_0$, 
  the first term accounts for
  intra-layer interactions, while the second one describes the
  interaction between layers. (The size of each layer
  is chosen such that only adjacent layers have non-zero interaction.
  It therefore depends on the details of the tight-binding approximation.)
  Finally the last term describes the coupling between the molecular
  region and each lead. Note that only the first layer $l_0$
  contributes to that term and that there is no interaction between
  different leads. These are only formal restrictions, as parts of
  each lead can be included into the molecular region.

  The determination of the transmission function involves two steps.
  First the conduction properties of 
  the isolated leads have to be calculated. 
  Thereby each lead will be decomposed into conducting 
  and non-conducting in- and out-going channels.
  These correspond to propagating and evanescent solutions
  moving in one of two possible directions respectively.
  In a second step, the channels are connected to each other
  via the molecular region. This is described by the scattering matrix
  and the transmission function
  is finally obtained by summing up the contribution from
  each channel.
  
  The calculation can be performed either using Green's function
  techniques\cite{Datta1995:Book} or
  equivalently\cite{FisherLee1981:PRB} using
  ESQC\cite{Sautet1988:PRB,Sautet1989:CPL}, which is a
  scattering-matrix approach.  We present the details of the
  calculation in the second scheme, as individual contributions from
  each channel to the transmission function can then be easily
  studied.
  
  \subsection{Bulk propagator}
  First we will restrict our attention to the semi-infinite lead
  Hamiltonians, which do not have to be identical.  The Hamiltonians
  of equation \eqref{eq:SystemHamiltonian} for one lead $H^d_{ll'}$
  are layer independent, if one assumes periodicity, i.e.\ $H^d_{ll} =
  H^d_{l_0l_0}$ and $H^d_{l,l+1} = H^d_{l_0,l_0+1}$.  Using Bloch's
  theorem one can reduce the infinite dimensional system of equations
  to an $N\times N$-matrix equation ($N$ being the number of orbital basis
  functions in one layer)
  \begin{equation}\label{eq:layerHamilton}
    \left(M(E)+h(E)e^{ik\Delta}+h^\dagger(E)e^{-ik\Delta}\right) 
    \gamma(k,E) = 0,
  \end{equation}
  with $M(E):=H_{l_0l_0}-ES_{l_0l_0}$,
  $h(E):=H_{l_0,l_0+1}-ES_{l_0,l_0+1}$, and $S_{ll'}$ is the overlap
  matrix between orbitals in layer $l$ and layer $l'$ for cases when
  one does not deal with an orthonormal basis set (otherwise
  $S_{ll'}=\text{Id}\cdot \delta_{ll'}$).  With $\Delta$ we denote the
  lattice spacing.  Defining $\lambda := e^{ik\Delta}$ one can easily
  see that equation \eqref{eq:layerHamilton} is an $N\times N$ quadratic
  eigenvalue equation.  It can be transformed into a $2N\times 2N$ linear
  eigenvalue equation:
  \begin{align}\label{eq:layerEigenEquation}
    P(E)\left(\begin{array}{c}\gamma_1\\\gamma_2\end{array}\right)
      &= \lambda(E) \left(\begin{array}{c}\gamma_1\\\gamma_2\end{array}\right),\\
    P&:=\left[\begin{array}{cc}
          0 & 1\\
          -h^{-1}h^\dagger & -h^{-1}M\\
      \end{array}\right]
  \end{align}
  (where we have dropped the energy dependency for notational ease).
  This layer-to-layer propagator $P$ reduces the problem of finding
  solutions for the entire isolated lead Hamiltonian to specifying the
  wave function coefficients at two adjacent layers $\gamma_l$ and
  $\gamma_{l+1}$ only.  This is due to the fact that given these
  coefficients any other pair of coefficients $\gamma_{l+2k},
  \gamma_{l+2k+1}$ can now be obtained via:
  \begin{equation}\label{eq:Propagator2}
    \left(
    \begin{array}{c}
      \gamma_{l+2k}\\
      \gamma_{l+2k+1}
    \end{array}
    \right) = P^k
    \left(
    \begin{array}{c}
      \gamma_l\\
      \gamma_{l+1}
    \end{array}
    \right).
  \end{equation}
  All possible solutions at energy $E$ can be decomposed into
  independent channels, by solving for the eigenvalues of equation
  \eqref{eq:layerEigenEquation}.  These eigenvalues come in pairs such
  that for each eigenvalue $\lambda_>$, there exists a corresponding
  eigenvalue $\lambda_<$ satisfying the relation $\lambda_> = 1/
  \lambda^*_<$, as can be seen by transposing equation
  \eqref{eq:layerHamilton}.  Eigenvalues with $\lvert\lambda\rvert\neq
  1$, i.e.\ complex $k$, belong to exponentially diverging solutions
  (see equation \eqref{eq:Propagator2}).  These are of course non
  physical, as long as the lead is infinite. In semi-infinite leads
  however (which we are dealing with), exponentially decaying
  coefficients at the boundary will contribute to the surface wave
  function and must not be neglected.
  
  \subsection{Current operator}
  The contribution from a single channel to the net current can not
  directly be seen from equation \eqref{eq:layerEigenEquation}.  It
  depends on the current density associated with a solution to the
  Schr\"{o}dinger equation $i\hbar\partial_t S\gamma=H\gamma$ and is
  obtained via the continuity equation.  The probability amplitude
  $|\gamma|^2$ for a stationary solution is constant in time
  \begin{equation}
    \frac{\partial}{\partial t}\gamma^\dagger S\gamma
    = \frac{i}{\hbar}\left(
         \gamma^\dagger H \gamma-
         \gamma^\dagger H^\dagger\gamma)\right)
    = 0,
  \end{equation}
  because $H$ and $S$ are hermitian. 
  For the probability amplitude 
  at all layers between $l_1$ and $l_2$ one therefore has
  \begin{align}\label{eq:CurrentOperator2}
\nonumber
    0 =&\frac{\partial}{\partial t}\sum_{l=l_1}^{l_2}(\gamma^\dagger_l\gamma^{\phantom x}_l)
\\\nonumber
      =&\quad\frac{i}{\hbar}\sum_{l=l_1}^{l_2}
          \gamma^\dagger_{l}(H-H)\gamma^{\phantom x}_{l}
\\
      =&\quad\frac{i}{\hbar}(\gamma^\dagger_{l_1-1}h(E)\gamma^{\phantom x}_{l_1} + 
                         \gamma^\dagger_{l_1+1}h^\dagger(E)\gamma^{\phantom x}_{l_1} -
                         \text{h.c.})
\\\nonumber
      &+ \frac{i}{\hbar}(\gamma^\dagger_{l_2-1}h(E)\gamma^{\phantom x}_{l_2} + 
                         \gamma^\dagger_{l_2+1}h^\dagger(E)\gamma^{\phantom x}_{l_2} -
                         \text{h.c.})
\\\nonumber 
    =&\quad
    \langle \gamma |l_2,l_2+1\rangle
      \frac{i}{\hbar}
      \left[\begin{array}{cc}
        0          & -h\\
        h^\dagger & 0
      \end{array}\right]
    \langle l_2, l_2+1| \gamma\rangle
\\\nonumber
    &- 
    \langle\gamma |l_1-1,l_1\rangle
      \frac{i}{\hbar}
      \left[\begin{array}{cc}
        0          & -h\\
        h^\dagger & 0
      \end{array}\right]
    \langle l_1-1, l_1| \gamma\rangle,
  \end{align}
  with the projectors $\langle l|\gamma\rangle :=\gamma^{\phantom
    x}_l$.  This gives rise to the definition of the current operator
  $W_l$ for layer $l$ as
  \begin{equation}\label{eq:currentOperator}
    W_l := 
    |l,l+1\rangle
    \frac{i}{\hbar}
    \left[\begin{array}{cc}
        0         & -h\\
        h^\dagger & 0
    \end{array}\right]
    \langle l, l+1|.
  \end{equation}
  Now let both $\varphi$ and $\vartheta $ be solutions at fixed energy
  $E$ with the eigenvalues $\lambda_1$ and $\lambda_2$ respectively.
  Because the expectation value for $W_l$ is layer independent
  (equation \ref{eq:CurrentOperator2}) one has:
  \begin{align}\label{eq:currentOperator2}
    \nonumber
    \langle \vartheta |W_l|\varphi \rangle  =& \langle \vartheta |W_{l+1}|\varphi \rangle\\
    =& \lambda_1^* \lambda_2 \langle \vartheta |W_l|\varphi \rangle.
  \end{align}
  This equation describes the connection between the current
  properties of a solution $\varphi$ and its eigenvalue $\lambda$.  We
  summarise the results of a detailed analysis of this equation, which
  is given in appendix \ref{ap:current}.  Each channel
  $|\varphi_i\rangle$ can be assigned a current value $v_i$, defined
  as
  \begin{equation}\label{eq:currentValue}
    v_i := \text{Im}\langle \varphi_i|W|\varphi_i\rangle,
  \end{equation}
  where we have used the layer independence of $W_l$ in simply writing $W$.
  
  Channels with eigenvalue modulus $|\lambda| \neq 1$, i.e.\ 
  evanescent waves have zero current value. They therefore do not
  contribute to the current. (Yet they are important at the surface,
  as already mentioned above.)
  
  Only channels with an eigenvalue of modulus one ($|\lambda|=1$)
  contribute to the current. The sign of $v_i$ determines the
  direction of charge transport.
  
  Therefore the eigenvalues can be sorted into incoming and outgoing
  solutions ($\lambda_>$ and $\lambda_<$ respectively), according to
  the following scheme: if $|\lambda| < 1$ it represents an incoming
  evanescent solution, if $|\lambda| > 1$ then it is outgoing
  evanescent. Only if $|\lambda| = 1$ it belongs to a propagating
  solution, which is incoming for $v>0$ and outgoing otherwise.
  
  We now define $\Lambda_>$ and $\Lambda_<$ as the two $N\times N$ diagonal
  matrices composed of all incoming and outgoing eigenvalues
  $\Lambda_\gtrless:=\text{diag}(\lambda_\gtrless^i)$.  The $2N\times
  2N$-matrix $U$, which diagonalises $P$:
  \begin{equation}
    U^{-1}PU =
    \left[
      \begin{array}{cc}
        \Lambda_> & 0   \\
        0   & \Lambda_<
      \end{array}
      \right],
  \end{equation}
  has the following quadratic block form:
  \begin{equation}
    \left[
      \begin{array}{cc}
        U_>    & U_<    \\
        U_>\Lambda_> & U_<\Lambda_<
      \end{array}
      \right].
  \end{equation}
  After this transformation into the diagonal basis of the propagator,
  we can easily obtain all physically relevant solutions of the
  infinite lead by specifying the amplitudes of all propagating waves
  at one lattice site.

  \subsection{Scattering matrix}
  Up to now, we have considered the isolated leads only. These are now
  assumed to be each coupled to the molecular defect region and
  thereby indirectly coupled to one another. We are interested in
  stationary solutions, which consist of an incoming propagating wave
  in one lead, being scattered among all the accessible outgoing
  channels (propagating and evanescent ones). This information is
  contained in the scattering matrix $S$
  \begin{equation}\label{eq:Smatrix}
    \left(\begin{array}{l}
      \mathcal{B}\\\mathcal{C}
    \end{array}\right)
    =\underbrace{
        \begin{bmatrix}
          s_{11}&s_{12}\\s_{21}&s_{22}
        \end{bmatrix}}_{=:S}
    \left(\begin{array}{l}
      \mathcal{A}\\\mathcal{D}
    \end{array}\right),
  \end{equation}
  which determines the wave amplitudes of all outgoing waves
  $\mathcal{B,C}$ given the incoming ones $\mathcal{A,D}$.
  \begin{figure}
    \begin{picture}(0,0)%
      \includegraphics{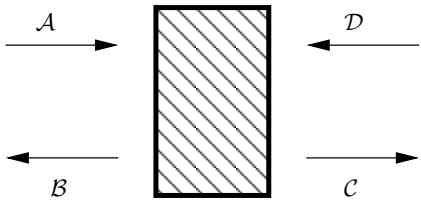}%
    \end{picture}%
    \setlength{\unitlength}{4144sp}%
    \begin{picture}(2499,1221)(439,-799)
      \put(631,254){$\mathcal{A}$}
      \put(721,-736){$\mathcal{B}$}
      \put(2476,-736){$\mathcal{C}$}
      \put(2476,254){$\mathcal{D}$}
    \end{picture}
    \caption{Incoming and outgoing wave amplitudes.}
  \end{figure}  
  In appendix \ref{ap:smatrix} it is shown, that $S$ is of the form
  \begin{equation}
    S = -M_{\text{in}}^{-1}\cdot M_{\text{out}}.
  \end{equation}
  
  It is important to notice, that the scattering matrix is always
  quadratic, because in each lead there are the same amount of
  incoming and outgoing channels.  This is opposed to the transfer
  matrix $T_\text{transf}$, which determines the amplitudes of in- and
  outgoing waves $\mathcal{C,D}$ in one lead given the in- and
  outgoing waves $\mathcal{A,B}$ of a second lead:
  \begin{equation}
    \left(\begin{array}{l}
      \mathcal{C}\\\mathcal{D}
    \end{array}\right)
    =T_\text{transf}
    \left(\begin{array}{l}
      \mathcal{A}\\\mathcal{B}
    \end{array}\right).
  \end{equation}
  This matrix is quadratic only if both leads have the same number of
  channels.  It then is of the form\cite{Rosh1992:PHD}
  \begin{equation}\label{eq:Transfermatrix}
    T_\text{transf} = 
      \begin{bmatrix}
      F&G^\dagger\\G&F^\dagger
      \end{bmatrix}
  \end{equation}
  and the relation to the scattering matrix is\cite{Rosh1992:PHD}
  \begin{equation}
    S = 
      \begin{bmatrix}
        -F^{\dagger(-1)}G & F^{\dagger(-1)}\\
        F^{-1}   & G^\dagger F^{\dagger(-1)}
      \end{bmatrix}.
  \end{equation}
  Methods calculating the scattering matrix via the transfer
  matrix\cite{Sautet1988:PRB} fail, if two types of leads are used,
  because $F$ is then no longer quadratic and can not be inverted.
  Therefore one commonly takes source and drain lead to be identically
  constituted.  But even in such cases, the method becomes numerically
  unstable, with increasing distance between the molecular region and
  one lead.  This is because the matrix elements of $F$ and $G$ (in
  equation \eqref{eq:Transfermatrix}) diverge exponentially, with
  increasing lead separation. Taking the inverse of $F$ is therefore a
  numerically critical procedure.  Both these problems are avoided by
  the direct calculation of the scattering matrix, which we present in
  appendix \ref{ap:smatrix}.  This calculation is well defined without
  any restrictions to the number of leads and their composition. That
  means that it is not necessary to restrict to identical leads.
  Furthermore it allows a numerically stable determination of the
  scattering matrix, even for weak coupling.

  \subsection{Transmission function}
  The transmission function is the sum over the contributions from
  each combination of incoming channel in the source lead to outgoing
  channel in the drain lead: $T(E) = \sum_{i,j}T_{j\gets i}$.  The
  relation between scattering matrix $S$ and these transmission
  function elements is
  \begin{equation}
    T_{j\gets i} = |(s_{21})_{j\gets i}|^2 \frac{v_j}{v_i},
  \end{equation}
  where $s_{21}$ is the lower left block of $S$ as defined in equation
  \eqref{eq:Smatrix}.  The weighting with velocity factors comes about
  because the scattering matrix $s$ does not relate current densities,
  but wave amplitudes.  The current densities are obtained from these
  wave amplitudes by multiplication with the corresponding velocity
  factor $v_j$.  The factor $v_i$ in the denominator normalises the
  transmission function to be exactly one for perfect transmission.
  If $v_i=0$ then $T_{j\gets i} = 0$, because incoming evanescent
  waves have zero amplitude at the surface.
  
  The total current is made up of the contribution from each
  $k$-state\cite{Datta1995:Book}:
  \begin{align*}
    I &= -\frac{e}{V}\sum_{k,\sigma} \sum_{i,j} v_i T_{j\gets i}(E_k) 
          \underbrace{\left(f(E_k-\mu_1)-f(E_k-\mu_2)\right)}_{=:\Delta f(E_k)}\\
      &= -\frac{e}{V}\sum_{k,\sigma} \sum_{i,j} \frac{\partial E_k}{\hbar \partial k} 
          T_{j\gets i}(E_k) \Delta f(E_k)\\
      &= -\frac{2e}{h} \int_{-\infty}^\infty {\rm d}E \sum_{i,j}T_{j\gets i}(E) \Delta f(E)\\
      &= -\frac{2e}{h} \int_{-\infty}^\infty T(E) \Delta f(E) {\rm d}E .
  \end{align*}
  Here the summation over $k$ has been transformed into an integral
  over $E$ with a factor of 2 accounting for spin $\sigma$.
  
  \section{Calculations for PDI}
  Low temperature experiments with PDI SAM's sandwiched between two
  metallic leads show several peaks in the
  $CV$-diagram.\cite{Lee2003:NanoLet,Dupraz2003:unpublished} The
  typical voltage differences of these peaks are in the range of
  $\Delta U\approx 0.2\text{V}$ (i.e.\ there are about 5 peaks within
  $U=0\text{V}$ and $U=1\text{V}$).
  The commonly adopted explanation 
  for the occurrence of such peaks is the following.
  Each molecular orbital that enters the 
  energy window, which is opened by the applied voltage, 
  enables resonant tunneling. 
  This increases the conductance and therefore results in a peak within
  the $CV$-diagram.
  
  Typically, the energy gap between molecular orbitals is in the range
  of $\Delta E\approx 1\text{eV}$. In other words, for applied
  voltages up to $U=1\text{V}$ there should be only a single
  accessible orbital per molecule, giving rise to only a single peak
  in the $CV$-diagram.  Therefore the following question arises: are
  there geometrical alignments of the molecules such that the
  additional peaks in the $CV$ diagram can also be explained by
  resonant tunneling through molecular orbitals?

  \subsection{Influence of changes in the molecular alignment 
    to the transmission spectrum} 
  During the device fabrication, the step under least experimental
  control is the adsorption of the molecules onto the leads.
  Therefore the exact geometrical alignment of the molecular SAM and,
  at least in the sandwich geometry, also the atomic shape of the top
  metallic lead, is not exactly known.  One therefore has to expect
  not only one specific but rather quite a variety of molecular
  alignments to be produced.  As one is interested in the conduction
  properties of the resulting device, it is important to understand
  the influence of each type of geometrical alignment to the
  transmission function.
  
  To this end, we have investigated three such possible alignments,
  which will be discussed separately.
  
  \begin{enumerate}
  \item 
      By adding metal-atoms on top of the molecular mono-layer in a random way,
      there might occur metallic clusters on top of the mono-layer.
      These affect the electronic configuration of the 
      molecules individually, 
      and might therefore have an influence to the transmission function. 
  \item 
      In a SAM experiment, there is not just one molecule, but rather
      a few hundred molecules involved.
      If the contribution to the transmission function 
      was different for each molecule, then
      $T(E)$ would change qualitatively,
      with a change in size of the mono-layer.
      We therefore analyse how the transmission functions depends
      on the number of molecules involved.
  \item
      If the molecular mono-layer is not strictly periodic,
      then there will be defects. 
      For example can the distance between two molecules be reduced, 
      such that inter-molecular bonds can be build. 
      Each of these defects will have a specific electronic structure
      and will therefore influence the transmission function.
  \end{enumerate}

  \subsubsection{Influence of metallic clusters}
  In the sandwich geometry, first the bottom metallic lead is created.
  Then the molecular mono-layer is adsorbed on top of it by
  self-assembly.  Finally the top metallic lead is build upon the
  molecular mono-layer.  The exact shape of neither metallic surface
  is known and may be anything but flat and regular.
  
\begin{figure}[h]
    \begin{minipage}{0.45\linewidth}
      \setlength{\unitlength}{\linewidth}
      \begin{picture}(0.98,0.72)%
        \put(0,0){%
          \includegraphics[width=\linewidth]
          {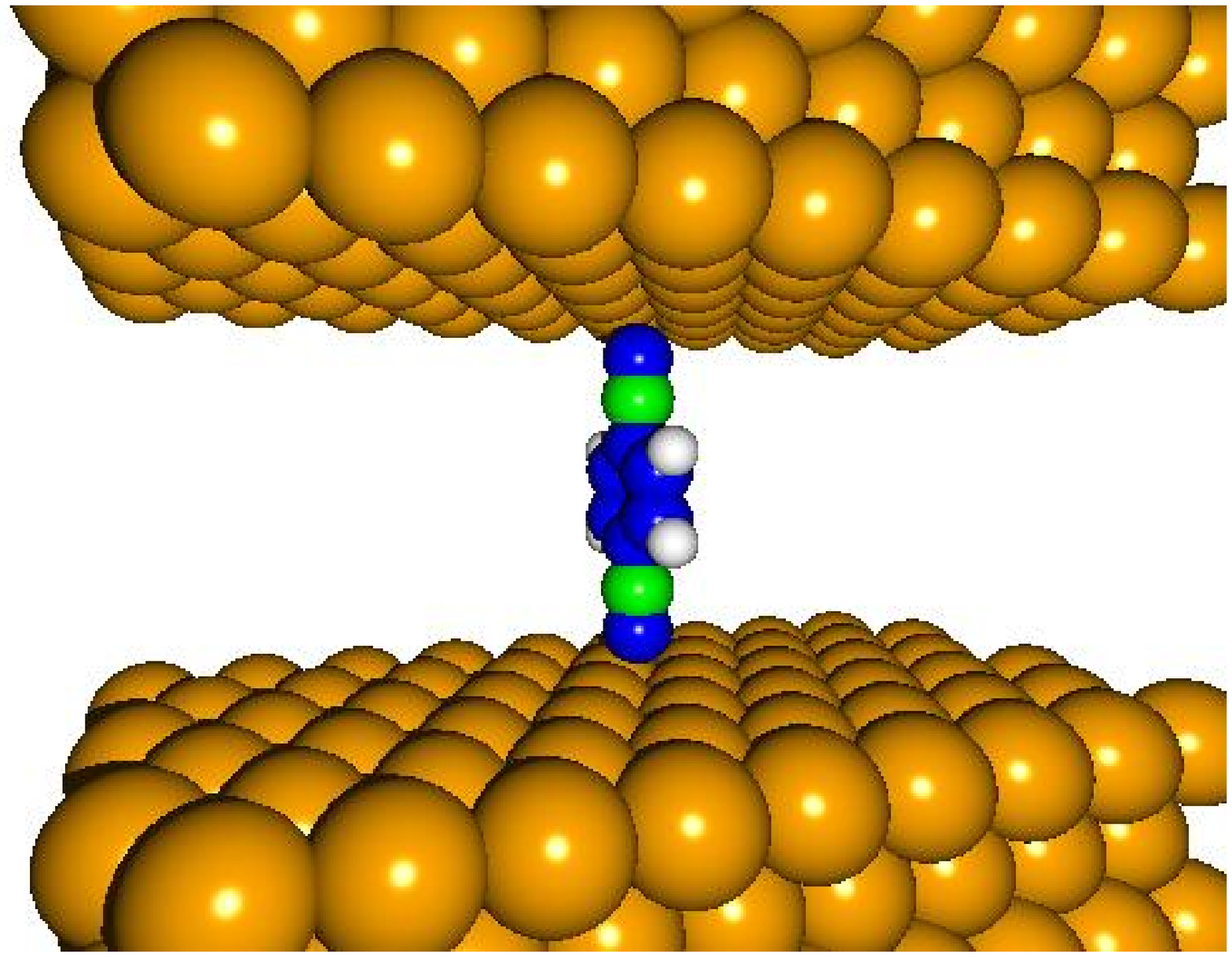}
        }
        \put(0,0.72){\makebox[0ex][r]{\raisebox{-2ex}{(a)}}}
      \end{picture}
    \end{minipage}\hfill
    \begin{minipage}{0.45\linewidth}
      \setlength{\unitlength}{\linewidth}
      \begin{picture}(0.98,0.72)%
        \put(0,0){%
          \includegraphics[width=\linewidth]
          {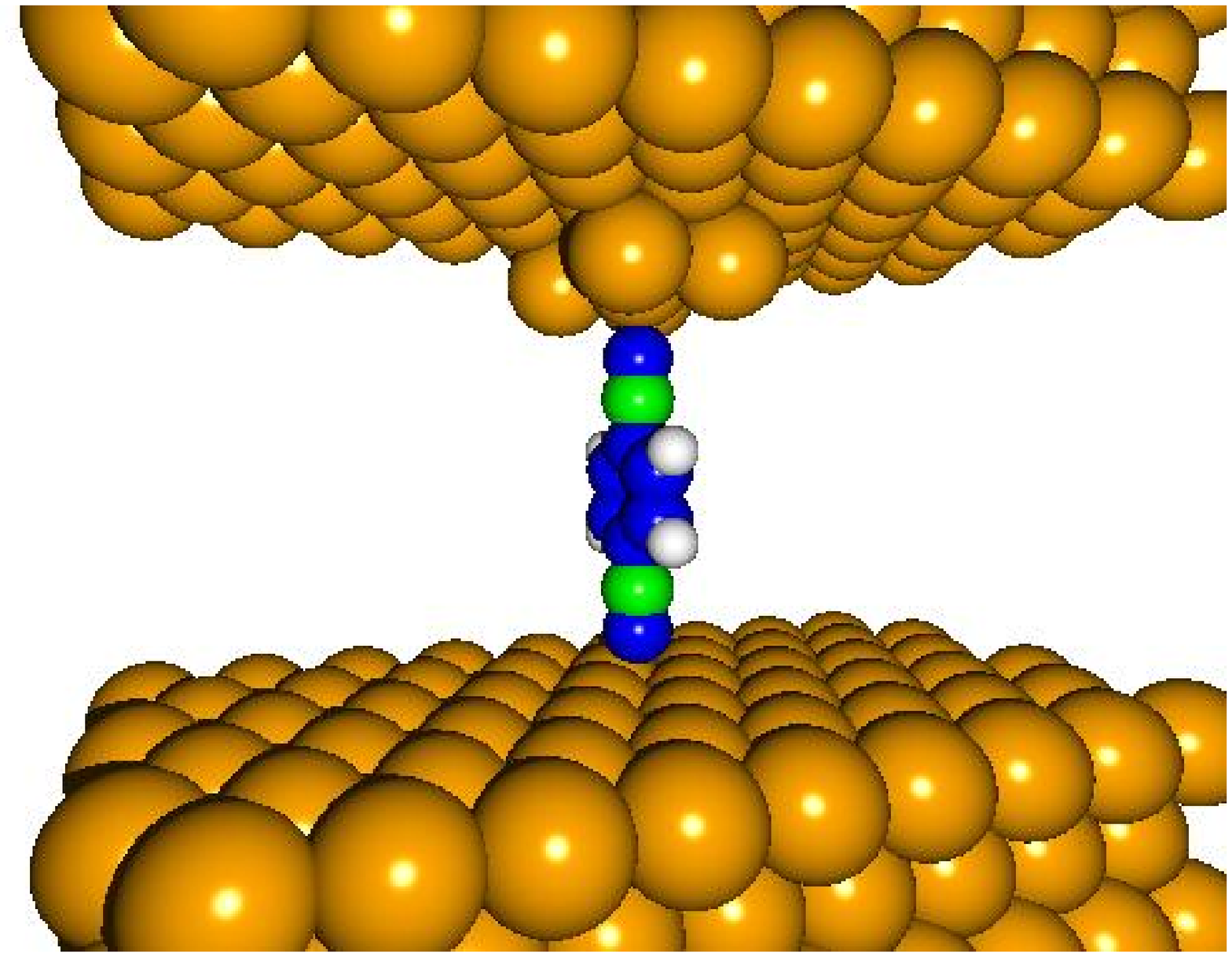}
        }
        \put(0,0.72){\makebox[0ex][r]{\raisebox{-2ex}{(b)}}}
      \end{picture}
    \end{minipage}
    \\[1ex]
    \centering
    \begin{minipage}{0.9\linewidth}
      \setlength{\unitlength}{\linewidth}
      \begin{picture}(0.98,0.98)%
        \put(0,0){%
          \includegraphics[width=\linewidth]
          {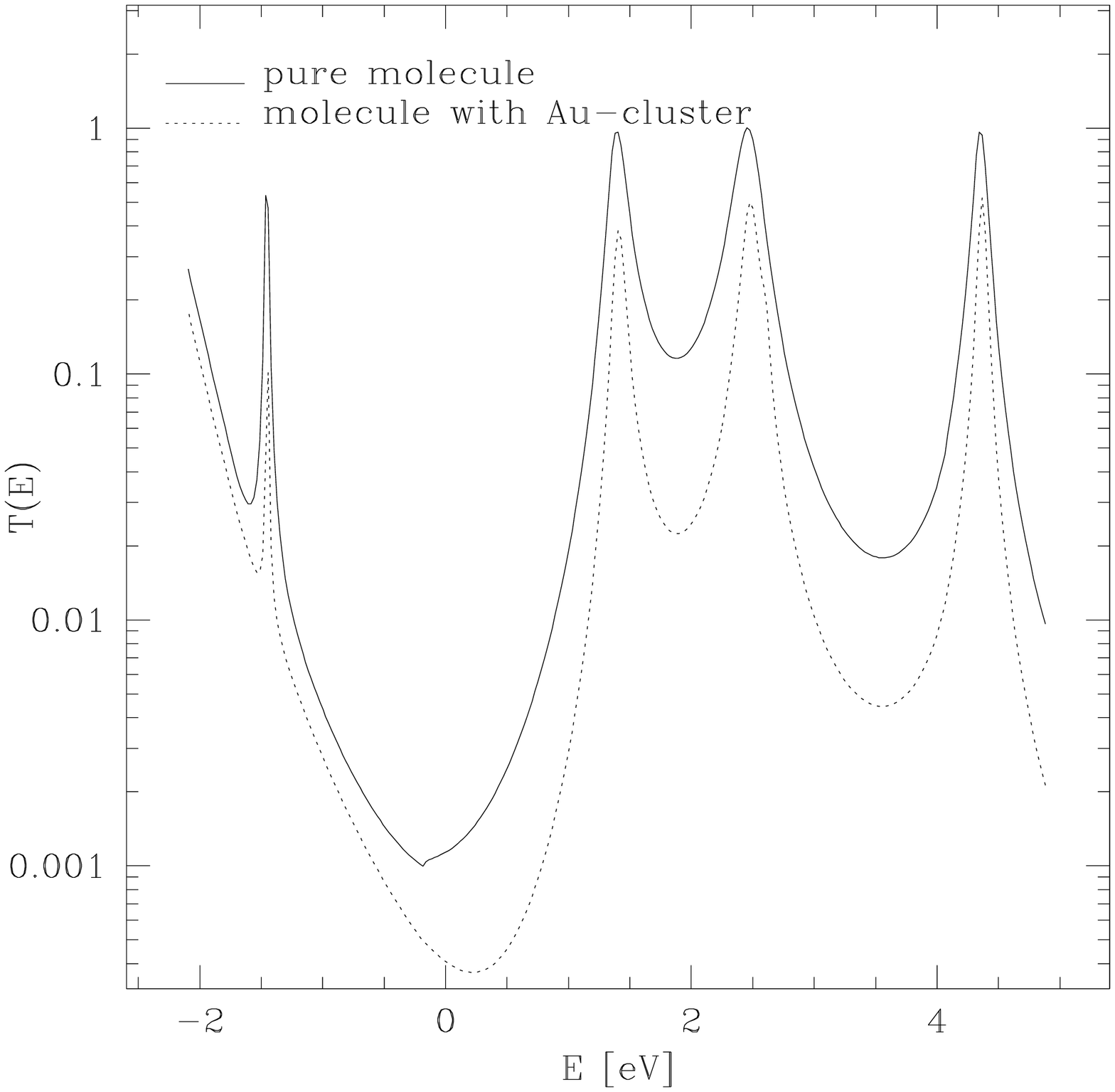}
        }
        \put(0,0.98){\makebox[0ex][l]{\raisebox{-4ex}{(c)}}}
      \end{picture}
    \end{minipage}
    \caption[Structure of Cluster]
          {\label{fig:singleAndCluster}
           (a) (Colour online) Structure for a molecule without cluster.
           (b) (Colour online) Sturcture for a molecule with 
           a gold cluster on top. 
           (c) Transmission function $T(E)$ for both structures.
           Energy scale is relative to
           the HOMO-LUMO gap, such that $E=0$ corresponds to the
           middle of the gap.
         } 
  \end{figure}

  It is likely that the surface atoms of the top metallic lead build
  up clusters on top of the molecular layer (as for example in
  Fig.\ \ref{fig:singleAndCluster} b).  Which influence
  do they have on the electronic configuration of the molecule they
  are in contact with? And do the clusters act as small molecules with
  new electronic levels?

  The influence of an Au cluster on the molecular electronic
  structure is twofold. First it introduces new electronic levels, and
  second the existing molecular electronic levels will be shifted, by
  an amount which depends on the strength of the coupling between
  cluster and molecule.
  
  The latter effect will be observed as a shifted peak in the
  transmission function, only if the coupling between cluster and
  molecule is different to the coupling between top electrode and
  molecule.  For clusters similar to the one shown in Fig.\ 
  \ref{fig:singleAndCluster} (b), this is however not the case.
  The energetic peak positions are identical, as can be seen
  in Fig.\ \ref{fig:singleAndCluster} (c).  
  
  Furthermore, there are no additional peaks, which one might have
  expected because of the additional electronic levels of the cluster.
  The explanation of their absence is the following: an electronic
  level gives rise to a peak in the transmission function only, if the
  corresponding orbital wave function overlaps with both the top and
  bottom electrode. The overlap with the electrode the cluster is
  attached to (say top electrode) is of course large. The overlap with
  the bottom electrode consists of two parts. The direct overlap and
  the indirect overlap via the molecule.  The direct overlap is
  negligible due to the large spatial separation.  The indirect
  overlap depends on the molecular orbital wave function. If the
  energy of the cluster level does not coincide with a molecular
  energy level, then there is no indirect overlap. Only if two levels
  coincide, the indirect coupling is large, but in that case, there
  already exists a transmission peak due to the molecule itself.
  
  Therefore if transmission is already suppressed by the molecule (at
  all off-resonant energies), it can either be further reduced by
  off-resonant tunneling through the cluster, or it can (at best) be
  left unchanged by resonant tunneling through the cluster. Under no
  circumstances can transmission, once suppressed by the molecule, be
  afterwards increased by the cluster.  This in turn means, that
  metallic clusters can not give rise to additional peaks in the
  transmission spectrum.

  \subsubsection{Mono-layer versus single molecule}
  \begin{figure}[h]
    \includegraphics[width=0.24\linewidth]
                    {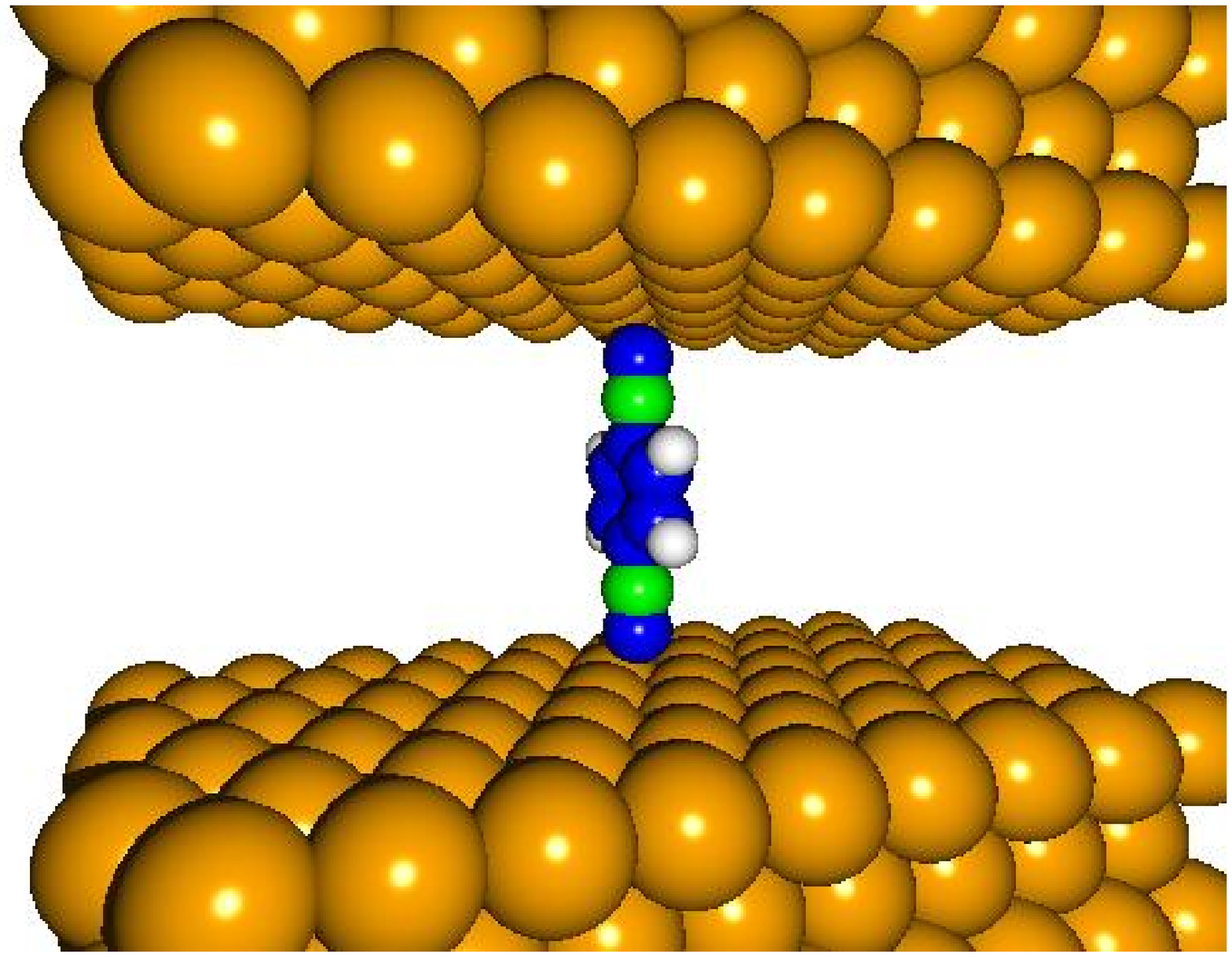}\hfill
    \includegraphics[width=0.24\linewidth]
                    {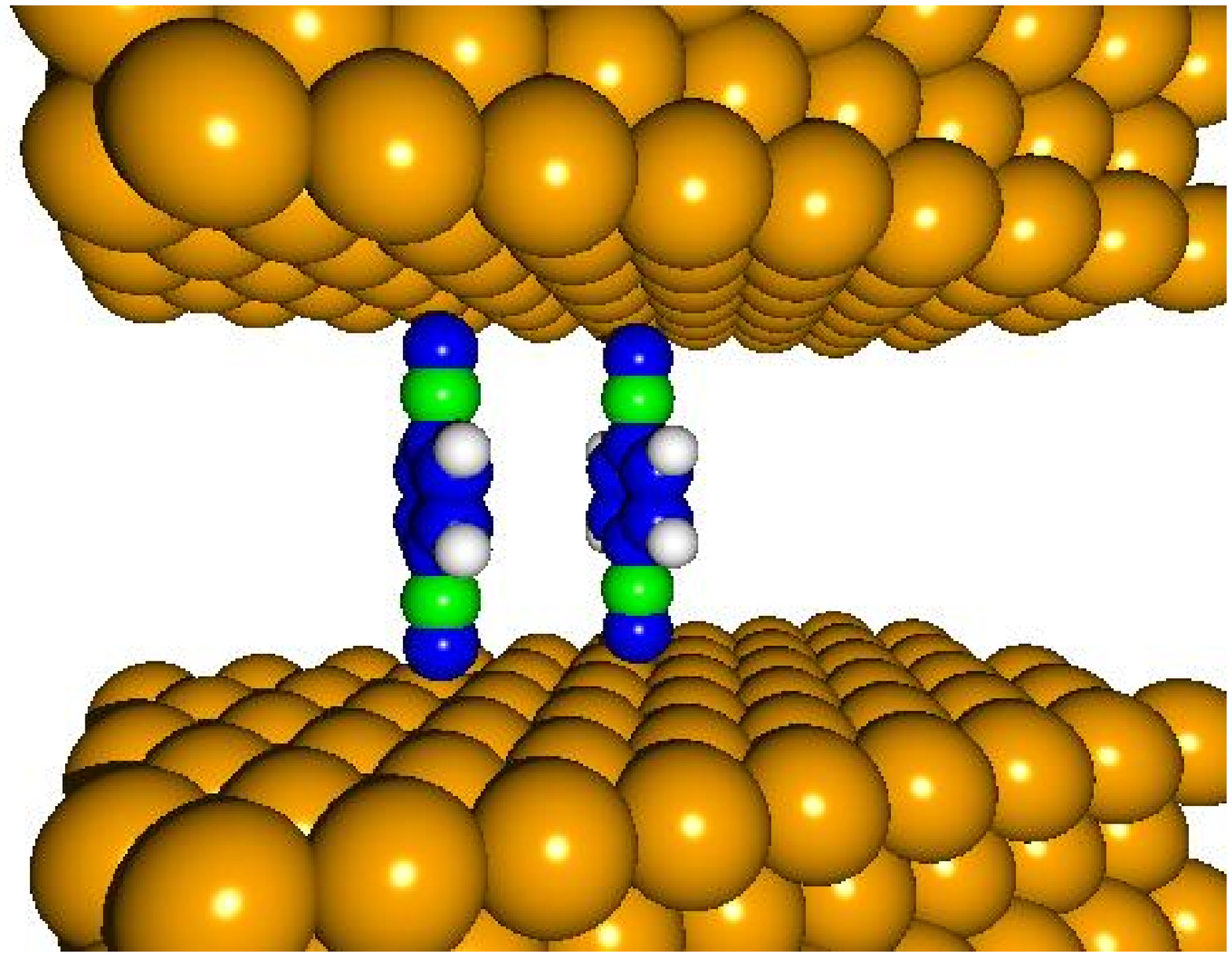}\hfill
    \includegraphics[width=0.24\linewidth]
                    {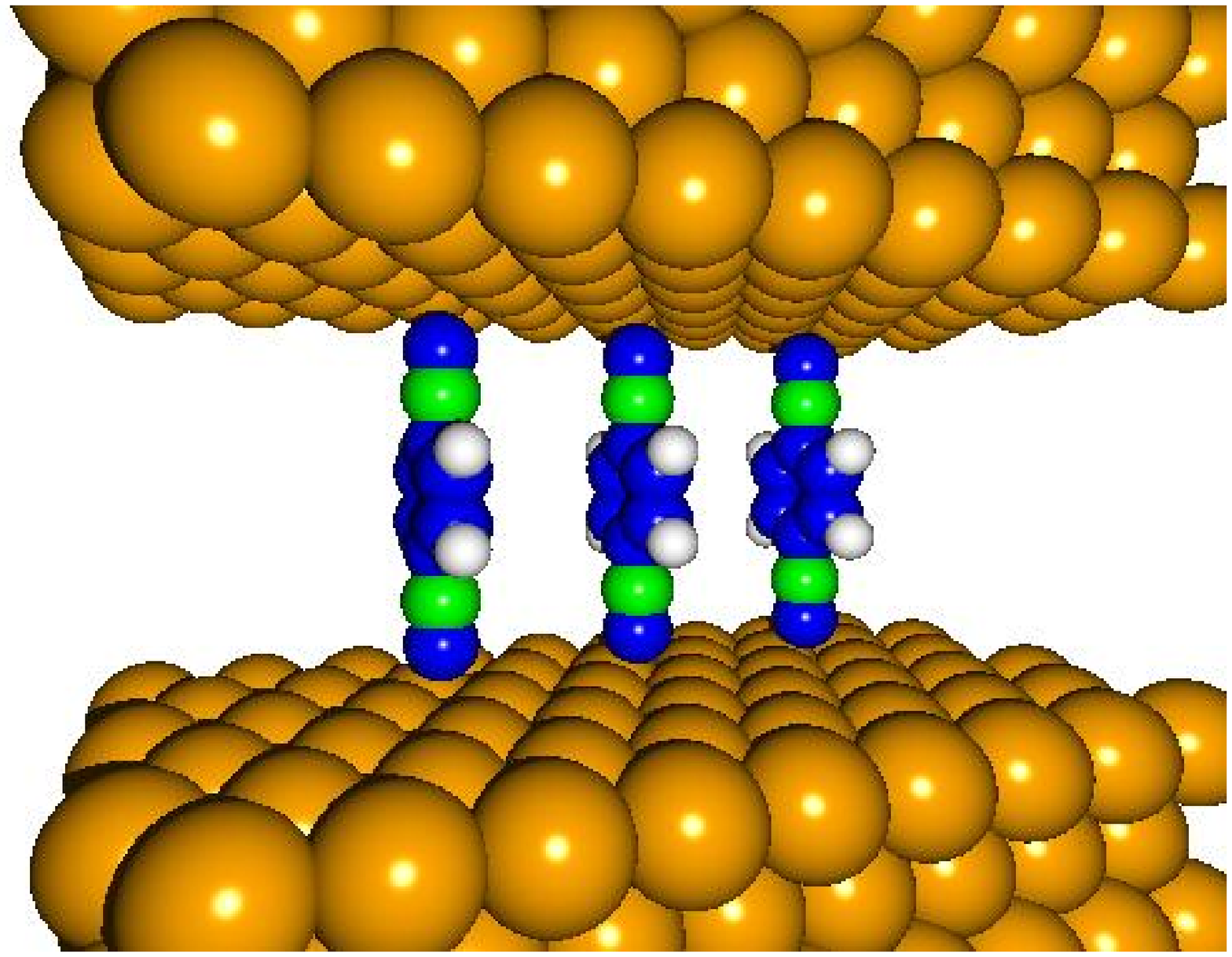}\hfill
    \includegraphics[width=0.24\linewidth]
                    {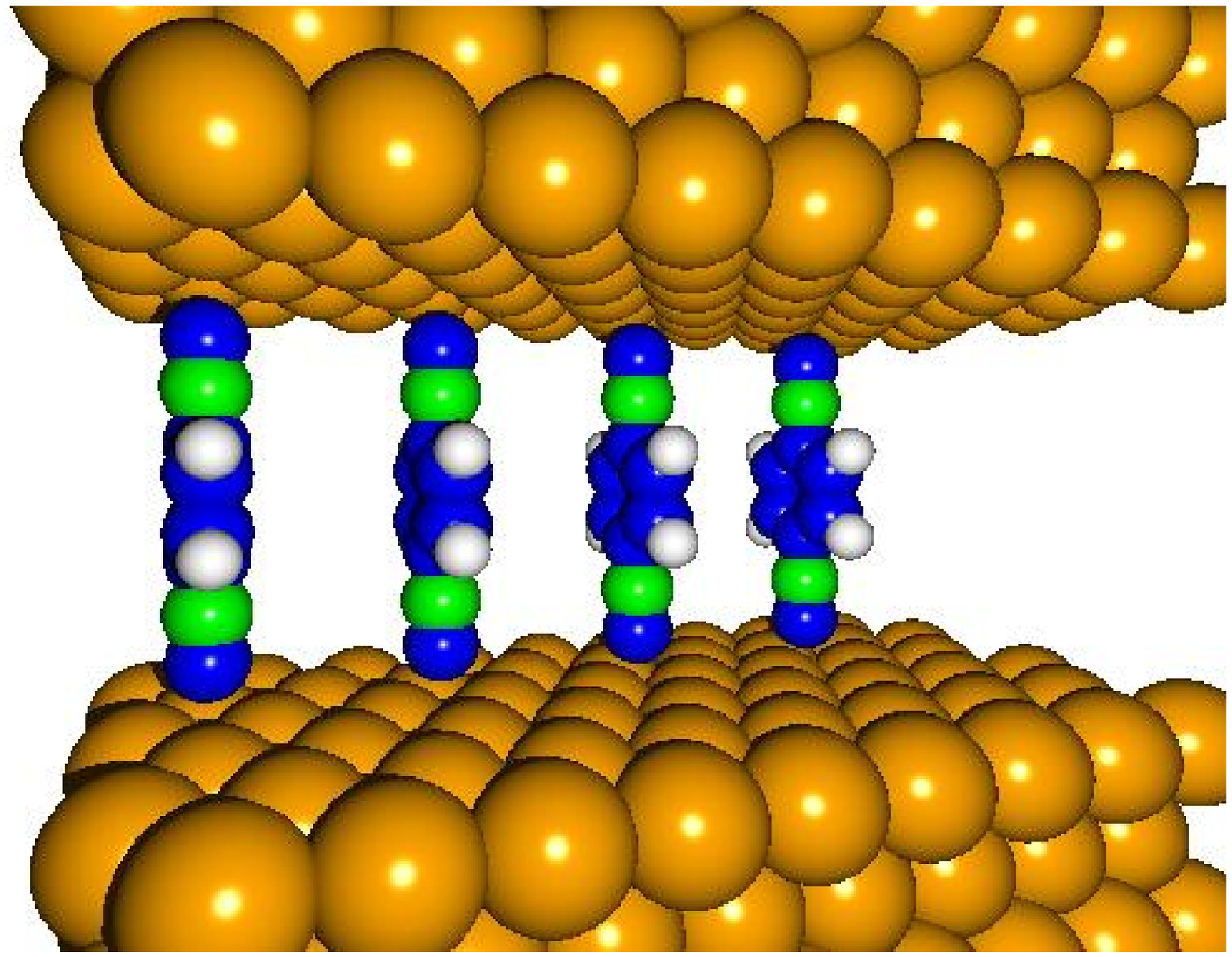}
    \caption[1,2,3 and four molecule structure]{
           \label{fig:oneToFourMolecules}
           (Colour online) The structure for one, two, three, and four molecules
           adsorbed within an Au-$9\times3$ super-cell. This setup was used to
           test the sum rule.}
  \end{figure}
  
  How do we expect the transmission function $T^i(E)$ for $i$
  periodically arranged molecules to look like?  As long as the
  inter-molecular interactions are small (compared to the
  intra-molecular ones) the molecular levels of each molecule will not
  be significantly changed.  Furthermore as the mono-layer consists of
  only one kind of molecule, all of them will have the same electronic
  structure.  Therefore we expect each molecule to contribute the same
  amount to the transmission function: $T^n(E) := \sum_i
  T^1(E)=nT^1(E)$, where $i$ runs over all $n$ adsorbed molecules.

  \begin{figure}[h]
    \centering
    \begin{minipage}{0.8\linewidth}
      \setlength{\unitlength}{0.98\linewidth}
      \begin{picture}(1.0,1.0)%
        \put(0,0){%
          \includegraphics[width=\linewidth]
          {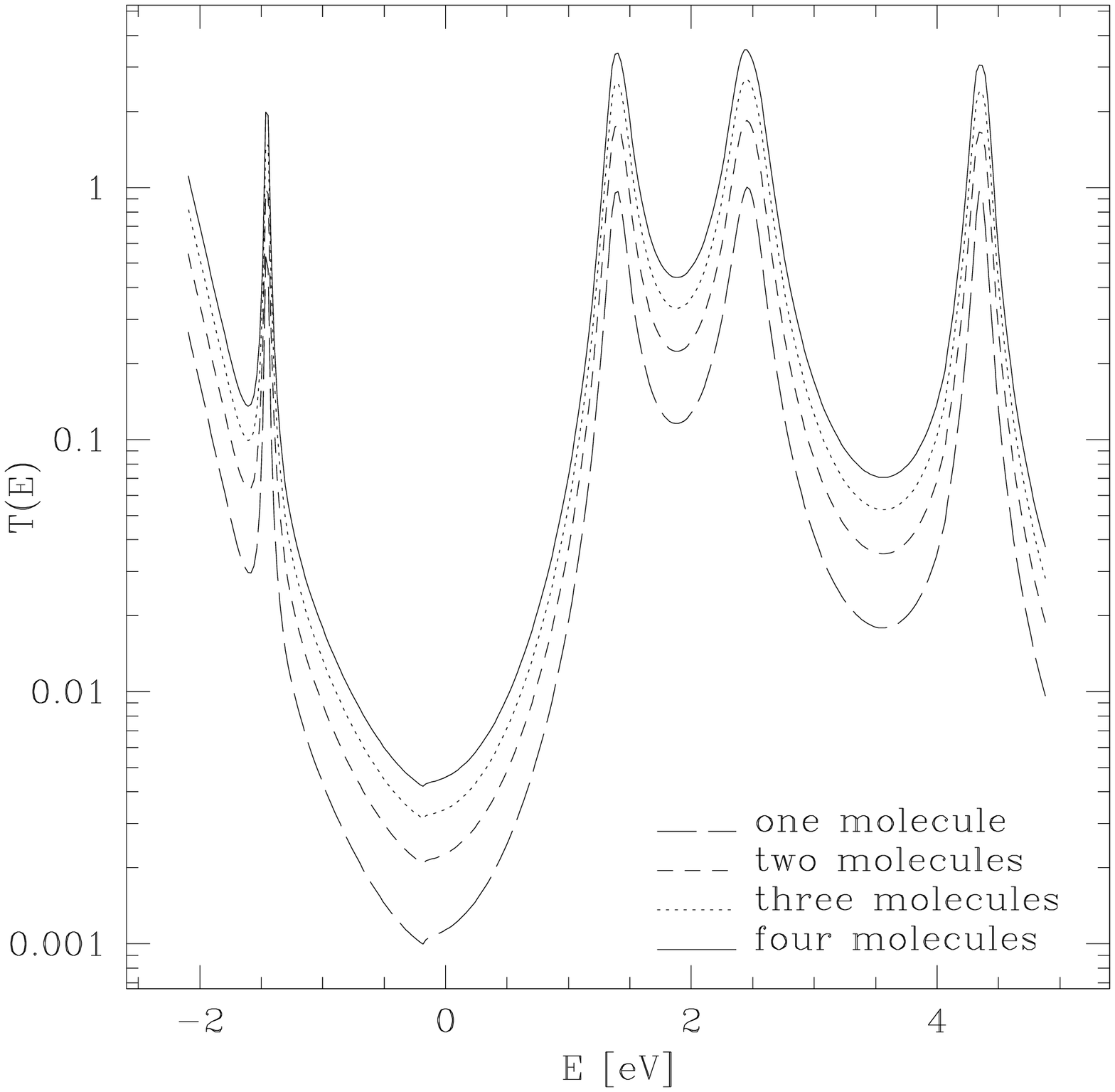}
        }
        \put(0,1.0){\makebox[0ex][l]{\raisebox{-4ex}{(a)}}}
      \end{picture}
    \end{minipage}

    \centering
    \begin{minipage}{0.8\linewidth}
      \setlength{\unitlength}{0.98\linewidth}
      \begin{picture}(1,1)%
        \put(0,0){%
          \includegraphics[width=\linewidth]
          {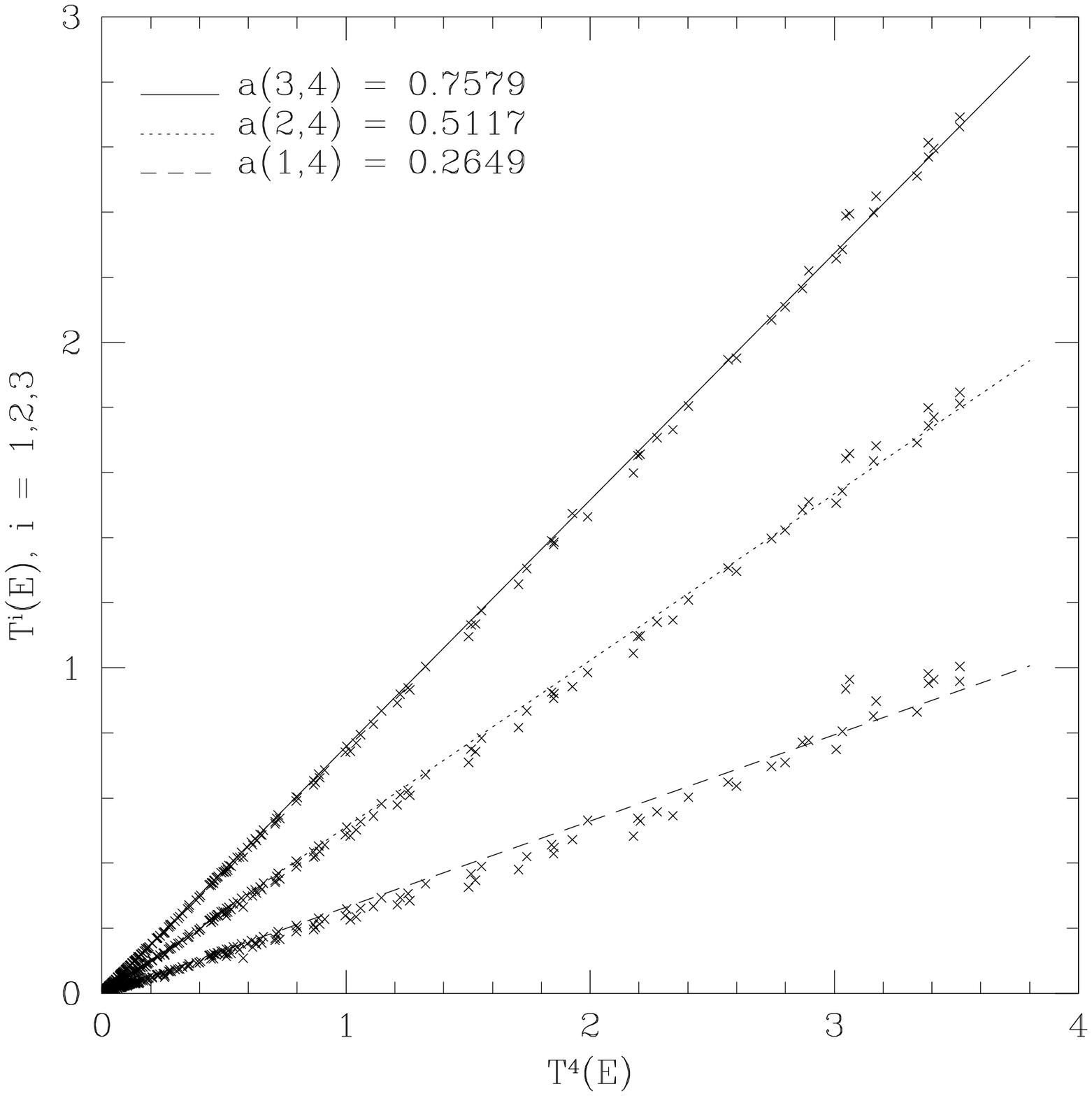}
        }
        \put(0,1){\makebox[0ex][l]{\raisebox{-4ex}{(b)}}}
      \end{picture}
    \end{minipage}
    \caption[Spectrum and fit for one to four molecules]{
        \label{fig:fourMoleculeSpectrum}
         (a) Transmission function for one, two, three, and
         four PDI molecules. 
         (b) When plotted against each other, the transmission
         functions reveal a linear relationship: 
         $T^i(E) = a(i,4)T^4(E)$ (bottom).}
  \end{figure}
  
  We calculated the transmission function for $n$=1 to 4 molecules
  within an Au super-cell of size $9\times3$ (the structures are shown
  in Fig.\ \ref{fig:oneToFourMolecules}).  The distance between the
  molecules is chosen to be a multiple of the closest Au-Au separation
  $a$ ($d=5.76\text{\AA}=2a$, with $a=2.88\text{\AA}$).  To our
  knowledge, the parameters of the PDI-SAM mono-layer have never been
  determined experimentally, which is why we have to assume the above
  values.  However STM studies\cite{Zeng2002:JChemPhys} and also
  theoretical calculations\cite{Yourdshahyan2002:JChemPhys} have been
  performed for alkanethiol mono-layers, and these parameters
  motivated our choice.
  
  Independent from the number of molecules present, the transmission
  functions have the same amount of peaks, at identical energetic
  positions (see Fig.\ \ref{fig:fourMoleculeSpectrum} a).  This result
  is also obtained for all larger distances of the molecules, where
  the inter-molecular interaction is even smaller.  Furthermore, the
  sum rule is indeed fulfilled, as shown in Fig.\ 
  \ref{fig:fourMoleculeSpectrum} (b). Each $T^i(E)$ is plotted against
  $T^4(E)$ for $i\in\{1,2,3\}$. The calculated transmission values
  (300 discrete values each) clearly show a linear correlation.  The
  straight lines are linear fits to the data, and their slope does
  very well agree with the theoretically expected value of
  $a(n,m)=n/m$.  The deviation is below $6\%$, as can be seen in table
  \ref{tab:fitparameters}, where we summarise all the fitted values
  for $T^n(E) =a(n,m) \cdot T^m(E)$.
  
  \begin{table}[h]
    \centering
    \begin{tabular}{llccc}\hline\hline
      $n$ & $m$ & $a(n,m)$ & $\frac{a}{n/m}-1$ & $\sigma$\\ 
      \hline
      1  & 2  & 0.5188   & 3.76\%            & 0.014\\ 
      1  & 3  & 0.3500   & 4.99\%            & 0.018\\
      1  & 4  & 0.2649   & 5.96\%            & 0.021\\ 
      2  & 3  & 0.6753   & 1.30\%            & 0.008\\
      2  & 4  & 0.5117   & 2.34\%            & 0.015\\
      3  & 4  & 0.7579   & 1.06\%            & 0.011\\
      \hline\hline
    \end{tabular}
    \caption[Fit-parameter $a$]
    {\label{tab:fitparameters}
      The fitted values for $a(n,m)$ together with their
      deviation from the theoretical value $a(n,m):=n/m$ and
      a measure for the quality of the fit $\sigma$, where
      $\sigma^2:=(N-1)^{-1}\sum\left(T^n(E_i)-a(n,m)T^m(E_i)\right)^2$ for
      $N=300$ discrete energy values $T(E_i)$.}
  \end{table}
  
  We conclude the following: a mono-layer, where the inter-molecular
  distance is large enough to not let inter-molecular interactions
  play a significant role, has the same number of distinct electronic
  levels as a single molecule. These levels are then highly
  degenerate.  A $CV$-diagram will therefore have the same number of
  peaks.  Only the net current will be increased by a factor $a(n,m)$
  compared to the single molecule case.
  
  The mere fact, that one deals with a mono-layer instead of a single
  molecule does not imply that the transmission function changes
  qualitatively.
  
  \subsubsection{Influence of molecular clusters}
  We have seen, that one does not observe additional peaks in the
  transmission function, as long as the inter-molecular influence is
  small.  And this is the case for distances which occur in typical
  SAM structures.\cite{Zeng2002:JChemPhys,Yourdshahyan2002:JChemPhys}
  We now investigate cases, where the molecular interactions are not
  negligible.  This occurs for example, when the periodic structure of
  the mono-layer is perturbed by an additional molecule, such that a
  molecular cluster is formed.  It is sufficient to study the
  transmission function of an isolated cluster only, because we have
  already seen that molecules in the periodic SAM arrangement do not
  influence each other.  The sum of the transmission function for the
  periodic SAM and the transmission function for the molecular cluster
  is, due to the sum rule, the total transmission function for defect
  and SAM.
  
\begin{figure}
    \centering
    \begin{minipage}{0.95\linewidth}
      \setlength{\unitlength}{0.98\linewidth}
      \begin{picture}(1.0,0.25)%
        \put(0,0){%
          \includegraphics[width=0.3\linewidth]
          {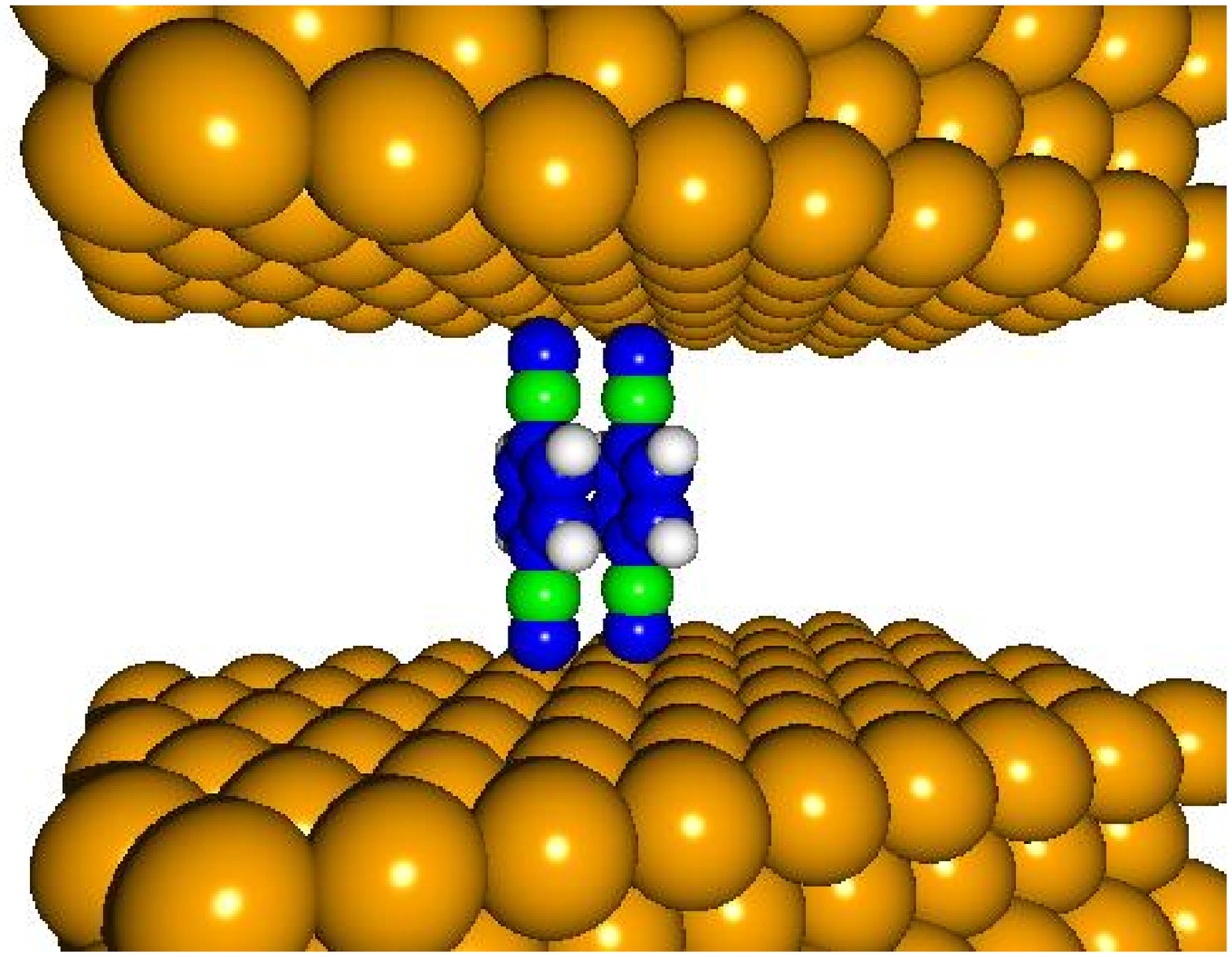}
        }
        \put(0.33,0){%
          \includegraphics[width=0.3\linewidth]
          {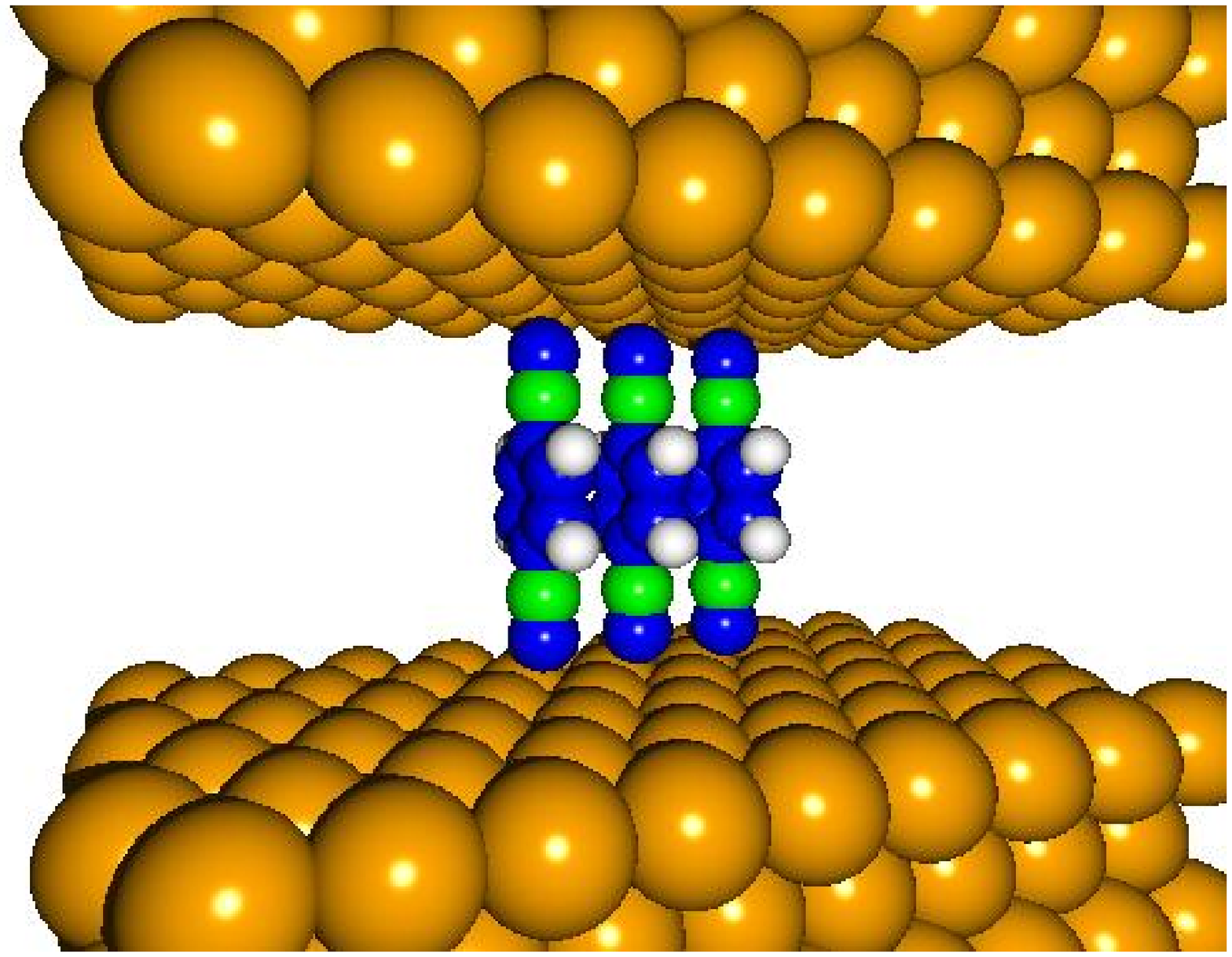}
        }
        \put(0.66,0){%
          \includegraphics[width=0.3\linewidth]
          {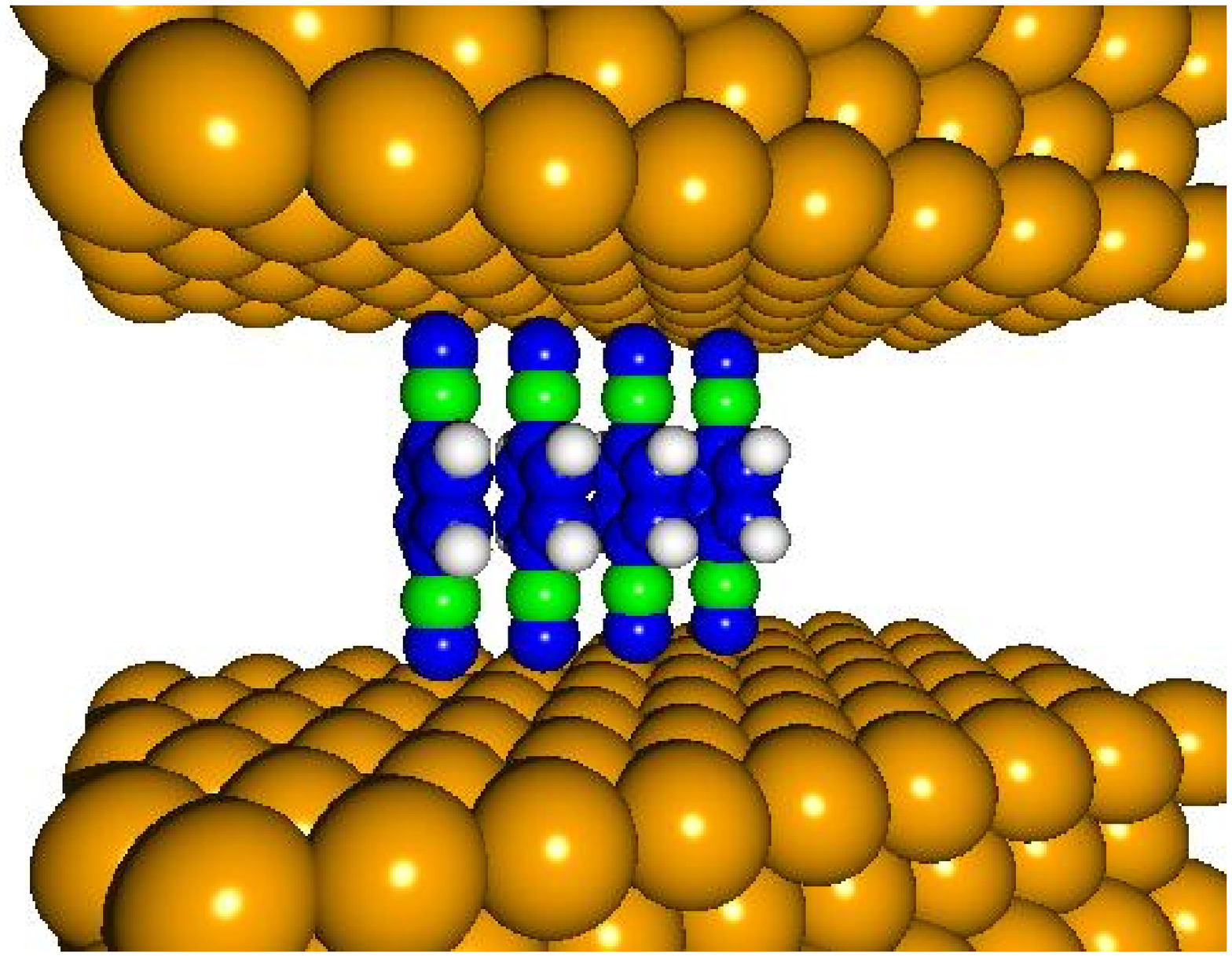}
        }
        \put(0,0.25){\makebox[0ex][r]{\raisebox{-4ex}{(a)}}}
      \end{picture}
    \end{minipage}

    \centering
    \begin{minipage}{0.8\linewidth}
      \setlength{\unitlength}{0.98\linewidth}
      \begin{picture}(1.0,1.0)%
        \put(0,0){%
          \includegraphics[width=\linewidth]
          {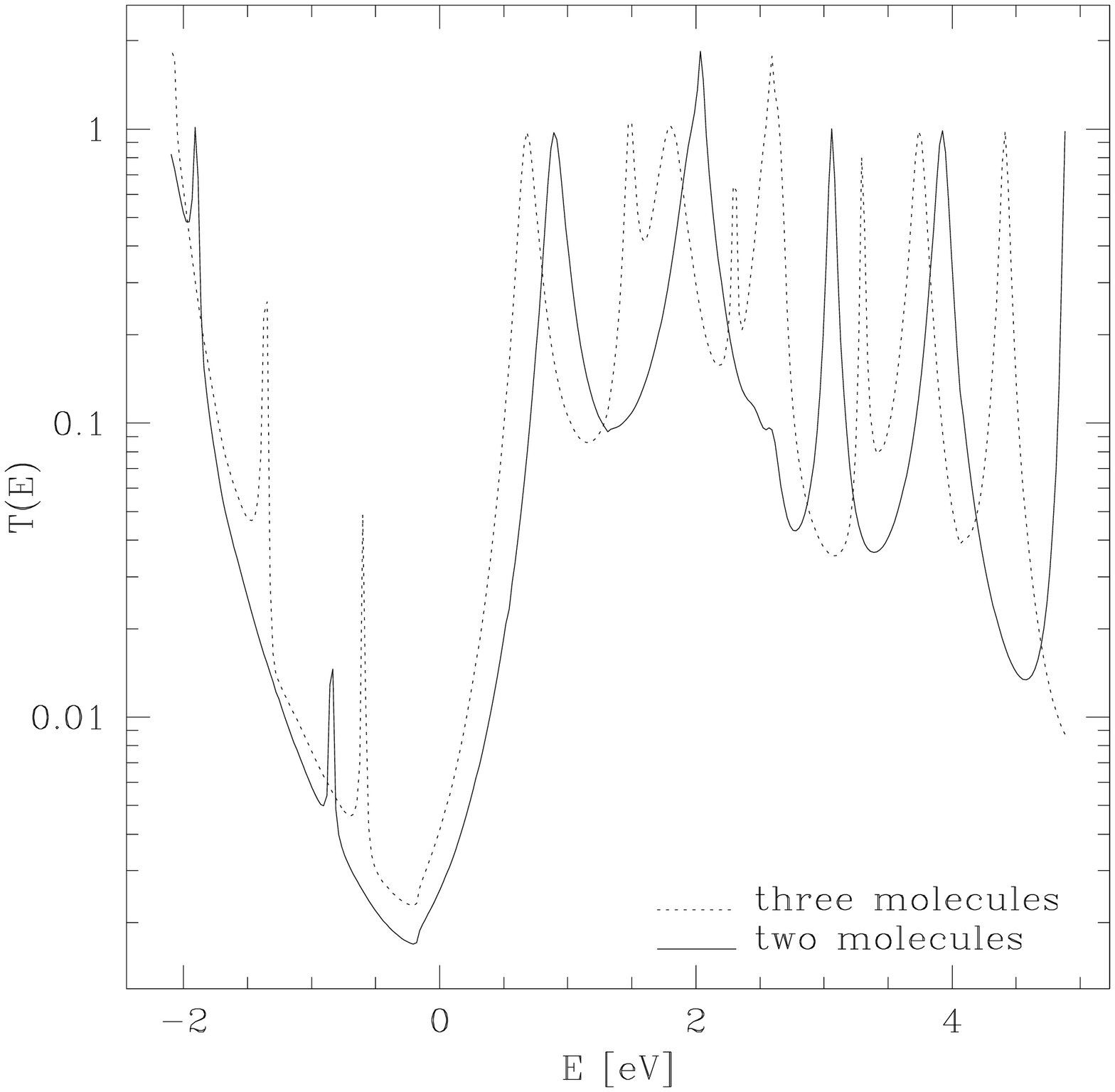}
        }
        \put(0,1.0){\makebox[0ex][l]{\raisebox{-4ex}{(b)}}}
      \end{picture}
    \end{minipage}

    \centering
    \begin{minipage}{0.8\linewidth}
      \setlength{\unitlength}{0.98\linewidth}
      \begin{picture}(1.0,1.0)%
        \put(0,0){%
          \includegraphics[width=\linewidth]
          {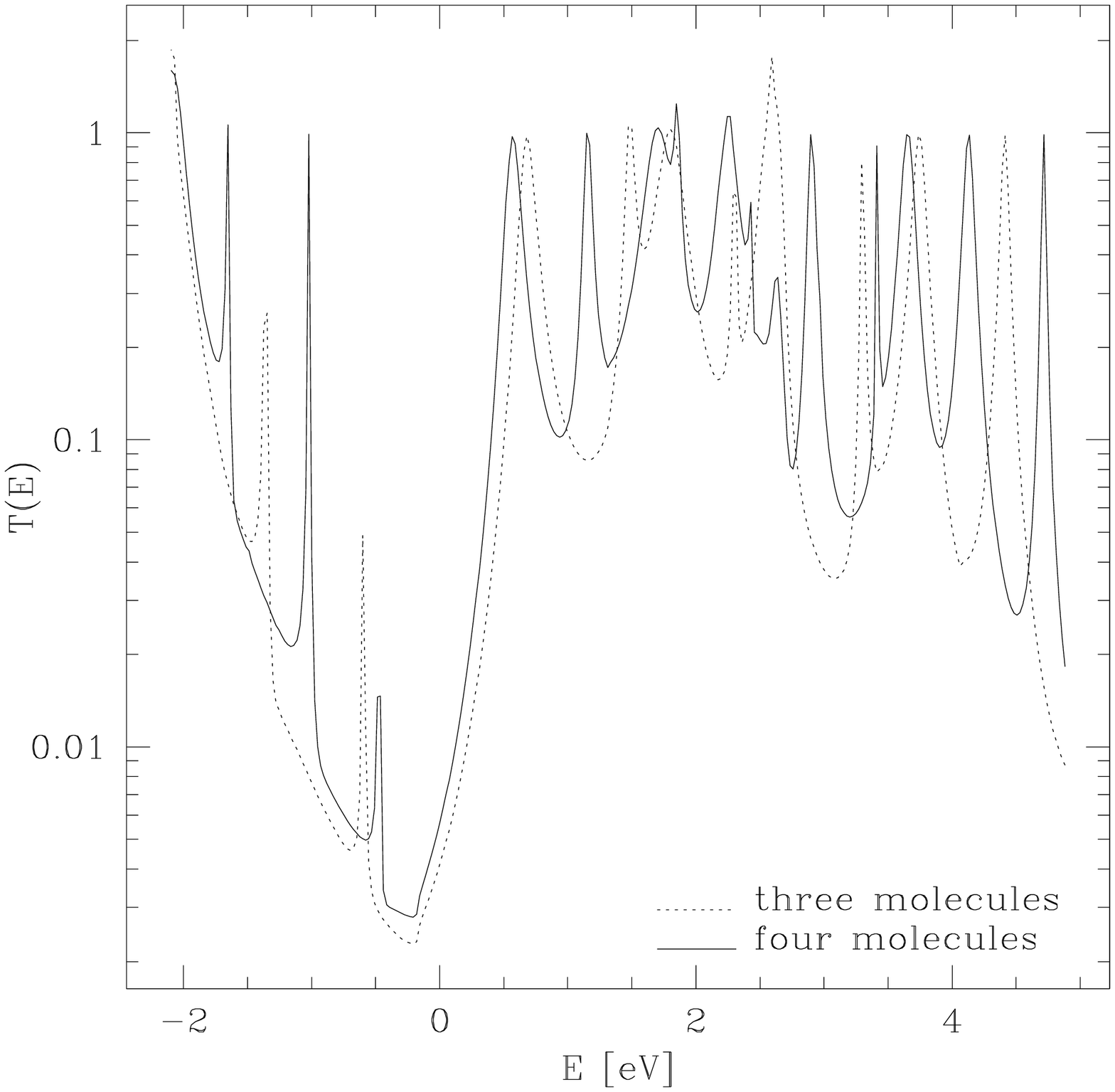}
        }
        \put(0,1.0){\makebox[0ex][l]{\raisebox{-4ex}{(c)}}}
      \end{picture}
    \end{minipage}
    \caption[Influence of molecular interaction]{\label{fig:molecularInteraction}
      (a) (Colour online) Two, three, and four molecules with a shorter inter-molecular 
      distance.         
      (b) The transmission functions for two and three molecules. 
      (c) The transmission functions for three and four molecules.
      In contrast to all previous cases,
      the peaks are shifted with respect to each other and
      there are also additional peaks. 
      These changes are due to the increase in 
      inter-molecular interaction, which alters the
      electronic levels.}
  \end{figure}

  We study the influence of a shorter distance between two, three, and
  four molecules on the transmission spectrum and relate it to the
  discrete energies of the isolated molecules.  The molecules are now
  separated by $d=2.88\text{\AA}$ which corresponds to the Au-Au atom
  spacing.  The atomic structure for this calculation is shown in
  Fig.\ \ref{fig:molecularInteraction} (a), the resulting transmission
  functions in Fig.\ \ref{fig:molecularInteraction} (b) and (c).
  
  By reducing the molecular separation from
  $d_1=2\cdot2.88\text{\AA}=5.76$\AA\ to $d_2=2.88$\AA, the transmission
  function qualitatively changes.  The number of peaks roughly
  doubles.  The new peak positions are different from the ones we have
  obtained in the previous calculations.  And this time, the peak
  positions do depend on the number of molecules involved.  This is an
  important point, because if there are several molecular clusters
  with different molecular distances, then they all give rise to peaks
  at different energy values.  The resulting transmission function is
  the sum of the individual functions and will thus contain far more
  peaks, than the transmission function for the non perturbed periodic
  layer.
  
  We now show that the new peaks are a result of the increase in
  molecular interaction due to the decrease in spatial separation.
  For non-interacting molecules, the molecular energies are identical
  and therefore degenerate.  An interaction between molecules breaks
  this degeneracy and therefore new energy levels occur.  By
  performing a diagonalisation of the molecular Hamiltonian (without
  leads) one can determine the levels of the molecular cluster.
  
\begin{figure}
    \includegraphics[height=0.8\linewidth]
    {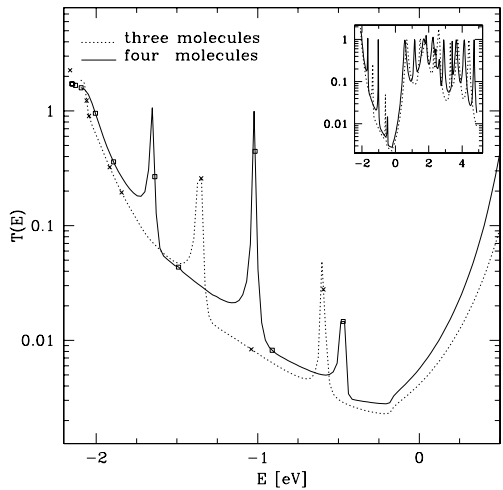}
    
    \caption[Influence of molecular interaction 2]
    {\label{fig:moleculesAndEnergies}
      Magnification of the transmission functions for three 
      and four closely spaced
      molecules.
      Additionally the discrete energy levels of the system without
      leads are plotted as points along the transmission function. 
      To each peak there belongs at least one discrete energy
      level. A detailed discussion is given in the main text.
      Inset: Transmission function (original scale) 
      for three and four closely
      spaced molecules (identical to Fig.\ 
      \ref{fig:molecularInteraction} c).
    }
  \end{figure}

  In Fig.\ \ref{fig:moleculesAndEnergies} we have again plotted the
  transmission function for three and four molecules, this time
  together with the discrete energy levels of the corresponding
  molecular cluster. The inset is identical to Fig.\ 
  \ref{fig:molecularInteraction} (c), while the plot itself is a
  magnification, to better resolve the discrete energy levels (which
  are shown as points along the transmission function).  Each of the
  transmission peaks is related to at least one discrete energy value.
  But not each energy value can be related to a peak in the
  transmission function.  Why is that?  The discrete energies can only
  give rise to new peaks in the transmission function, if they are not
  suppressed by a weak coupling to one of the leads. All levels which
  are not related to any peak belong to this category.  If the
  position of the peak is shifted away from a corresponding energy
  level, then this is due to the coupling between molecules and leads.
  This coupling is absent in the diagonalisation of the molecular
  Hamiltonian, but present in the calculation of the transmission
  function.
 
\begin{figure}
    \includegraphics[width=0.6\linewidth]
    {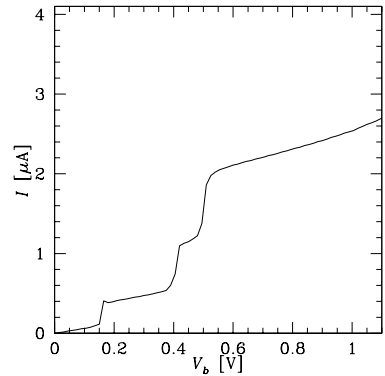}
    \caption{
      \label{fig:defectCurrent}$IV$--calculation for a molecular
      region containing all three molecular clusters shown in
      Fig.\ \ref{fig:molecularInteraction} (a). There are three
      distinct steps within the voltage range of 1V.
    }
  \end{figure}

  Finally we show, that the additional peak structure in the
  transmission function for a scenario with an increased
  inter-molecular interaction gives rise to a number of steps in the
  $IV$-curve. Figure \ref{fig:defectCurrent} contains an $IV$
  calculation for a molecular structure containing all three molecular
  clusters shown in Fig. \ref{fig:molecularInteraction} (a).  In this
  calculation, the bias voltage $V_b$ enters as a shift of the Fermi
  levels for source and drain lead: $\mu_1=\mu_2+eV_b$.  The molecular
  energy has been set to $E_m=\mu_1-\delta E_m-\eta eV$, where
  $\delta E_m$ is the zero bias displacement of
  the molecular levels and $\eta = 0.5$, because of the symmetric
  coupling to the leads.
  
  Compared to the
  experiments\cite{Lee2003:NanoLet,Dupraz2003:unpublished} the number
  of steps in the $IV$--curve is well reproduced by our calculation.
  The obtained current is at least one order of magnitude larger than
  the experimental values.\cite{Lee2003:NanoLet} This is a phenomenon
  common to all theoretical methods based on the Landauer 
  formula.\cite{Emberly2001:PRB,Bauschlicher2003:CPL} 
  A satisfactory explanation for this discrepancy as well as for
  the broad range of experimentally observed current values has not
  yet been found.  

  \section{Discussion}
  We have shown, that the peak structure of the transmission function
  is robust against changes in the number of adsorbed molecules, as
  long as the distance between molecules is considerably large
  ($d\gtrsim 6$\AA).  And also does the exact shape of the top
  metallic lead not influence the qualitative structure of the
  transmission function. Only if the distance between molecules
  becomes so small, that inter-molecular interactions are no longer
  negligible (which is below 6\AA\ in our case), does the transmission
  function undergo a qualitative change. Namely an additional peak
  structure occurs.
  
  How does this finding compare to the experimental data?  As we have
  pointed out in section \ref{sec:experiment}, only in devices using
  molecules with two cyanide end-groups a more or less random peak
  structure was observed in the $CV$
  characteristic\cite{Lee2003:NanoLet,Dupraz2003:unpublished}.  In
  other devices, molecules with at least one thiol end-group are
  typically used. These show significantly less peak structure.
  
  We therefore give the following interpretation: The thiol end-group
  is known to bind strongly to Au atoms.  It is therefore likely, that
  thiol-based mono-layers stably adsorb to gold leads. Resulting
  periodic structures are then robust against distortions. The
  conductance of such structures is proportional to the corresponding
  single molecule conductance, i.e.\ the number of molecules involved
  changes the absolute value of the current only, not the peak
  structure.
  
  The random like peak structure in devices, made up of cyanide based
  molecules suggests, that there are a some molecular clusters present
  in the mono-layer. These clusters might occur, because the binding
  of a cyanide end-group to Au is considerably weaker compared to that
  of a thiol end-group, and weaker binding results in a less robust
  periodic structure.
  
  \section*{Acknowledgements}
  We would like to thank Xavier Bouju, as well as Udo Beierlein for
  helpful discussions.

  \appendix
  \section{Connection between eigen- and current-values}\label{ap:current}
  The current properties of each channel can be related to the
  corresponding eigenvalue.  We start from equation
  \eqref{eq:currentOperator2}:
  \begin{align*}
    \langle \psi|W_j|\phi\rangle  =& \langle \psi |W_{j+1}|\phi\rangle\\
    =& \lambda_1^* \lambda_2 \langle \psi|W_j|\phi\rangle.
  \end{align*}
  Let's first consider $|\psi\rangle = |\phi\rangle$, i.e.\ $\lambda_1
  = \lambda_2$, i.e.\ $\langle \psi|W_j|\phi\rangle=|\lambda|\langle
  \psi|W_j|\phi\rangle$.  For each channel with eigenvalue $|\lambda|
  \neq 1$ one then must have $\langle \psi|W_j|\psi\rangle=0$, i.e.\ 
  this channel does not itself carry any current.  This is consistent
  with our terminology of an evanescent wave.  If, however,
  $|\lambda_i|=1$, then $\langle\psi|W_j|\psi\rangle$ is purely
  imaginary, because $W_j$ is an anti-hermitian operator. We can
  therefore define the velocity of a propagating wave to be
  $v_i:=\text{Im}\langle\psi|W_j|\psi\rangle$.
  
  Now we consider the case of two different solutions $|\psi\rangle
  \neq |\phi\rangle$ and define $v_{1,2}:=\langle
  \psi|W_j|\phi\rangle$.  If their eigenvalues do not satisfy
  $\lambda_1 \lambda_2^* = 1$, then the current between these two
  solutions is zero $v_{1,2}=0$.  So let's assume $\lambda_1 = 1/
  \lambda_2^*$. Because if $|\lambda_1|>1$ then $|\lambda_2|<1$, a
  current can flow between an evanescent left going wave and an
  evanescent right going wave. But if we restrict ourselves to
  solutions with finite amplitudes in a semi-infinite lead, then
  either the left or right going wave amplitude must be zero.
  Therefore evanescent waves do neither carry a current themselves nor
  do they exchange current with other channels, that is they do not at
  all contribute to the net current.
  
  Finally we are left with the case $\lambda_1=1/ \lambda_2*$, with
  $|\lambda_1|=|\lambda_2|=1$.  This is equivalent to
  $\lambda_1=\lambda_2$, i.e.\ the case of degenerate eigenvalues.
  Therefore propagating waves to degenerate eigenvalues do exchange
  current. That in turn means, that the current of a superposition of
  two such waves does not necessarily equal the sum of the two
  individual currents, which is problematic as we want to express the
  total current as a sum of independent channels.  However, the
  propagating and evanescent waves were obtained by diagonalising the
  propagator $P$. This transformation is unique up to rotations in
  every degenerate eigenvalue subspace. Because $W$ is anti-hermitian
  we can diagonalise these subspaces and the resulting diagonal
  elements will be purely imaginary.  So the net current may be
  written as a summation over all the individual contributions of
  propagating channels, only if these subspace rotations are
  performed.
  
  Summarising we have shown that the transformation $U$ diagonalising
  the propagator $P$ (i.e\ $U^{-1}PU$) can be chosen such that the
  transformation $U^\dagger W U$ of the current operator is diagonal
  in the subspace of propagating waves with purely imaginary diagonal
  elements.  All the other diagonal entries are zero and the only
  non-zero non-diagonal elements belong to evanescent waves in
  opposite directions.

  \section{Calculation of the scattering matrix\label{ap:smatrix}}
  The part of the Hamiltonian containing the molecular region and its
  coupling to the leads can be written as
  \begin{equation}\label{eq:scatHamilton}
    (H-ES)|\psi\rangle =\left[\begin{array}{ccccc}
        h_1 & M_1 & 0   & 0   & \tau_1^\dagger \\
        0   & 0   & h_2 & M_2 & \tau_2^\dagger \\
        0   & \tau_1 & 0   & \tau_2 & M_0         \\
      \end{array}\right] |\psi\rangle = 0.
  \end{equation}
  (Using this order for the coefficients it is straight forward to
  extend all formulas to the general case of more than two leads.) The
  indices 1 and 2 indicate source and drain lead surface layers, while
  the index 0 is used for the molecular region.  $\tau_{1,2}$ are the
  coupling matrices from source/drain to the molecules.
  
  We now transform into the basis of incoming and outgoing channels,
  i.e. we apply
  \begin{equation*}
    U=\left[\begin{array}{ccc}
        U^1 & 0   & 0\\
        0   & U^2 & 0\\
        0   & 0   & 1\\
        \end{array}\right],
    \quad\text{with }
    U^i=\left[\begin{array}{cc}
        U^i_>      & U^i_<    \\
        U^i_>\Lambda^i_> &U^i_<\Lambda^i_<\\
        \end{array}\right]
    \end{equation*}
  from the right to equation \eqref{eq:scatHamilton}:
  \begin{equation}\label{eq:channelScatHamilton}
    (H-ES)U=
    \left[\begin{array}{ccccc}
        A^1_> & A^1_< & 0     & 0     & \tau_1^\dagger\\
        0     & 0     & A^2_> & A^2_< & \tau_1^\dagger\\
        B^1_> & B^1_< & B^2_> & B^2_< & M_0\\
    \end{array}\right],
  \end{equation}
  with 
  \begin{align*}
    A^i_\gtrless &= h_iU^i_\gtrless+M_iU^i_\gtrless\Lambda^i_\gtrless,
    \quad\text{and}\\
    B^i_\gtrless &= \tau_iU^i_\gtrless\Lambda^i_\gtrless.\\
  \end{align*}
  The first and third column act on the surface layer of the incoming
  channels, the second and fourth act on outgoing ones, while the
  fifth column, acting on the molecular region, remains unchanged.
  
  The scattering matrix expresses the outgoing channel amplitudes in
  terms of the incoming ones. Therefore we split the matrix of
  equation \eqref{eq:channelScatHamilton} into two parts, one
  containing the outgoing columns, the other one containing the
  incoming ones as well as the molecular column:
  \begin{align*}
    M_\text{out} &:=
    \left[
      \begin{array}{ccc}
        A^1_< & 0     & \tau_1^\dagger\\
        0     & A^2_< & \tau_2^\dagger\\
        B^1_< & B^2_< & M_0\\
      \end{array}
    \right]\quad\text{and}\\
    M_\text{in} &:=
    \left[
      \begin{array}{cc}
        A^1_> & 0    \\
        0     & A^2_>\\
        B^1_> & B^2_>\\
      \end{array}
    \right].
  \end{align*}
  The first matrix $M_{\text{out}}$ is a square matrix and by
  inverting it, we obtain the scattering matrix
  \begin{equation}
    s = - M_{\text{out}}^{-1}\cdot M_{\text{in}}.
  \end{equation}

\end{document}